\documentclass[%
 reprint,
 amsmath,amssymb,
 aps,
]{revtex4-2}
\usepackage[utf8]{inputenc}

\usepackage{graphicx}
\usepackage{dcolumn}
\usepackage{xcolor}
\usepackage{bm}

\begin{document}

\preprint{}

\title{Ab initio quasi-harmonic thermoelasticity, piezoelectricity, and thermoelectricity of polar solids at finite temperature and pressure: Application to wurtzite ZnO}

\author{Xuejun Gong}
 \affiliation{School of Physics and Technology, Xinjiang University, Urumqi, Xinjiang 830046, China}
\email{xgong@sissa.it}
\email{xuejun.gong@xju.edu.cn}
 
\author{Andrea Dal Corso}
\affiliation{International School for Advanced Studies (SISSA), Via Bonomea 265, 34136, Trieste, Italy}
 \affiliation{IOM - CNR, Via Bonomea 265, 34136, Trieste, Italy}
\email{dalcorso@sissa.it}


\date{\today}

\begin{abstract}
We generalize a previously established ab initio approach—originally developed for hexagonal close-packed (hcp) metals—to accommodate solids with both internal and external degrees of freedom. This extension enables the thermodynamic and thermoelastic characterization of insulators, including those with non-vanishing piezoelectric and pyroelectric tensors. Utilizing Density Functional Theory (DFT) and Density Functional Perturbation Theory (DFPT) within the quasi-harmonic approximation, we derive the pressure and temperature dependence of these properties. Specifically, we investigate internal degrees of freedom using two distinct frameworks: the Zero Static Internal Stress Approximation (ZSISA) and Full Free Energy Minimization (FFEM). We then compare these approximations by computing internal and external thermal expansions, as well as temperature-dependent piezoelectric and pyroelectric tensors. Finally, we demonstrate the generalized formalism by calculating the thermodynamic properties of wurtzite ZnO across a broad range of pressures and temperatures.
\end{abstract}

\maketitle


\section{\label{sec:level1}Introduction}

A comprehensive thermodynamic description of insulators requires a precise determination of the pressure and temperature dependence of their thermoelastic, piezoelectric, and pyroelectric properties \cite{nye_physical_1985}. Furthermore, many technologically significant insulators possess internal degrees of freedom and the estimate of their internal thermal expansion represents a further challenge.

The computation of temperature-dependent elastic constants (TDECs) is currently supported by a well-established theoretical framework and validated workflows. These methodologies enable the evaluation of TDECs within both the quasi-static (QSA) and quasi-harmonic approximations (QHA) \cite{erba_combining_2014, rostami_anisotropic_2025, malica2021thesis, gong2024thesis}. Previous studies on various metallic systems have demonstrated that the QHA—despite its higher computational cost—consistently yields superior agreement with experimental data compared to the QSA. Consequently, the QHA is considered essential for achieving high-fidelity predictive results \cite{malica2021thesis, gong2024thesis}.

The theoretical foundation for calculating piezoelectric and pyroelectric tensors has been rigorously established over the past three decades, primarily through the Berry phase formulation of macroscopic polarization in solids \cite{king-smith_theory_1993, resta_macroscopic_1994}. However, only recently has the literature emphasized the necessity of advancing beyond the zero static internal stress approximation (ZSISA) to accurately model the temperature dependence of internal degrees of freedom \cite{liu_internal_2018, masuki_full_2023}. A recent example (within ZSISA) of thermodynamic calculations in insulators is provided in Ref. \cite{mathis_ab_2024}, which reports on the temperature-dependent piezoelastic properties of ferroelectric materials and their impact on elastic constants using advanced ab initio techniques.

Wurtzite ZnO--a prototypical piezoelectric and pyroelectric material-- has served as a primary benchmark for the Berry phase theory of polarization due to its significant electromechanical coupling properties \cite{dal_corso_ab_1994, hill_first-principles_2000, wu_systematic_2005, masuki_full_2023, rostami_anisotropic_2025, liu_internal_2018}. Its widespread study in the literature stems from its technological relevance in transducers and sensors, where the accurate modeling of structural responses to external fields is paramount.

Experimentally, the elastic constants, compliances, thermal expansion coefficients, and piezoelectric tensors of ZnO are well-characterized at ambient conditions \cite{bateman_elastic_1962, kobiakov_elastic_1980, ibach_thermal_1969}. Specifically, at zero pressure, the temperature dependence of certain elastic constants ($C_{33}$, $C_{55}$) and compliances ($S_{11}$, $S_{12}$, $S_{55}$) has been documented up to $800$ K~\cite{kobiakov_elastic_1980}. It is important to note that the former were measured at constant electric displacement ({\bf D}), while the latter were determined under a constant electric field (${\bf E}$). Furthermore, data for electromechanical coupling coefficients and static dielectric constants exist up to $800$ K \cite{kobiakov_elastic_1980}, while pyroelectric tensor measurements are available up to $400$ K \cite{ibach_thermal_1969}. Despite these contributions, significant gaps remain: the temperature dependence of several elastic constant components remains unmeasured, and while the pressure dependence of elastic constants has been explored at room temperature \cite{sarasamak_pressure-dependent_2010}, data characterizing the simultaneous effects of high pressure and temperature are unavailable. Regarding structural stability, the wurtzite phase of ZnO is known to persist at room temperature until approximately $90$ kbar, beyond which it undergoes a phase transition to a rocksalt structure \cite{decremps_pressure-induced_2001, jaffe_hartree-fock_1993, serrano_pressure_2004}.

Pioneering ab initio calculations of the piezoelectric components ($e_{31}$ and $e_{33}$) and elastic constants of ZnO were first reported in Refs.~\cite{dal_corso_ab_1994} and \cite{shein_elastic_2007}, respectively. While Hill and Waghmare \cite{hill_first-principles_2000} subsequently explored the temperature and stress dependence of the piezoelectric tensor, their thermodynamic model was restricted to zone-center phonon frequencies, thereby neglecting the full vibrational density of states. The first determination of the $e_{15}$ component followed in Ref. \cite{catti_full_2003}.
A more comprehensive suite of properties at $0$ K and $0$ kbar-- including elastic constants at both constant electric field and displacement, piezoelectric tensors, dielectric constants, and Born effective charges--was established by Wu et al.~\cite{wu_systematic_2005}. The pressure dependence of these parameters was later investigated in Ref. \cite{marana_piezoelectric_2017}. Regarding temperature effects, Wang et al. \cite{wang_thermodynamic_2014} computed the elastic constants within the quasi-static approximation (QSA). Most recently, Rostami et al. \cite{rostami_anisotropic_2025} reported the thermal expansion and elastic constants of ZnO utilizing the QHA, though their treatment of internal degrees of freedom was limited to the ZSISA.

The pyroelectric response of ZnO was theoretically characterized by Liu and Pantelides~\cite{liu_mechanisms_2018}, while the necessity of advancing internal thermal expansion models beyond the ZSISA framework was rigorously established in Ref. \cite{liu_internal_2018}. More recently, Masuki et al. \cite{masuki_full_2023} presented a comprehensive calculation of both thermal expansion and pyroelectric coefficients. Their approach utilized a full minimization of the Helmholtz free energy to determine internal degrees of freedom, representing a significant step toward the complete thermodynamic description of wurtzite structures.

In this work, we generalize a previously developed ab initio framework for the elastic constants of hcp metals \cite{malica_quasi-harmonic_2020, malica_quasi-harmonic_2021, gong_high-temperature_2024, gong_high-pressure_2025} to encompass insulating materials, including those with piezoelectric and pyroelectric properties. This generalized workflow is applied to ZnO to evaluate its thermodynamic and piezoelectric responses across a broad range of temperatures and pressures.

All calculations are performed within a consistent computational framework: the plane-wave method utilizing norm-conserving pseudopotentials and the PBEsol exchange-correlation functional \cite{pbesol}. We provide a comprehensive characterization of ZnO, including internal and external thermal expansion, heat capacity, and bulk moduli. The temperature-dependent elastic constants (TDECs) are evaluated within the QHA using the ZSISA scheme with both constant ${\bf E}$ and constant ${\bf D}$ conditions. For piezoelectric properties, the clamped-ion contributions and Born effective charges are calculated within the QSA, while the strain derivatives of the internal parameters are evaluated using both ZSISA and the Full Free Energy Minimization (FFEM)~\cite{gong_high-temperature_2024,gong_high-pressure_2025} Finally, we derive the pyroelectric tensor from the temperature-dependent internal parameters and compare our FFEM results with existing literature.

To maintain a concise narrative, well-documented properties are provided in the Supplementary Material,~\cite{supplemental} while the main text focuses on the novel aspects of the generalized formalism and the high-pressure/high-temperature regimes.

The remainder of this paper is organized as follows. In Section \ref{sec:level2}, we present the governing equations of the generalized workflow and detail the implementation of the FFEM approach. Section \ref{sec:level3} specifies the technical parameters and computational settings employed in our calculations. The results for ZnO are discussed in Section \ref{sec:level4}, where we evaluate the generalized formalism against previous theoretical studies and available experimental data. Finally, Section \ref{sec:level5} provides concluding remarks and outlines potential future perspectives for this work.

\begin{figure}
\centering
\includegraphics[width=0.8\linewidth]{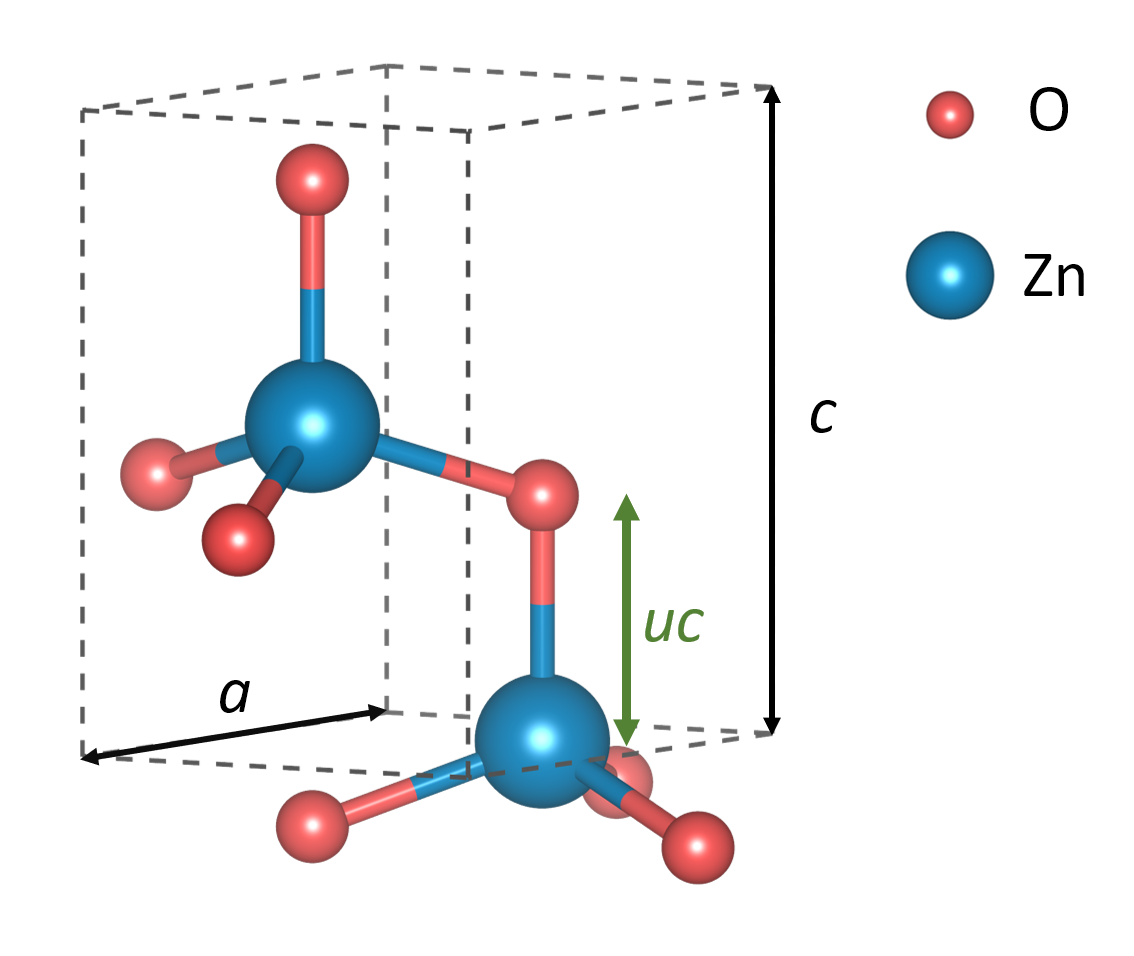}
\caption{The wurtzite crystal structure of ZnO, illustrating the internal and external parameters utilized in this study. The external lattice parameters are defined by the cell dimensions $a$ and $c$, while the internal parameter $u$ characterizes the relative position of the anion and cation sublattices along the $[0001]$ axis. (Rendered using VESTA 3~\cite{momma_vesta_2011})}
\label{fig:wurtzite_struc}
\end{figure}

\begin{figure}
\centering
\includegraphics[width=\linewidth]{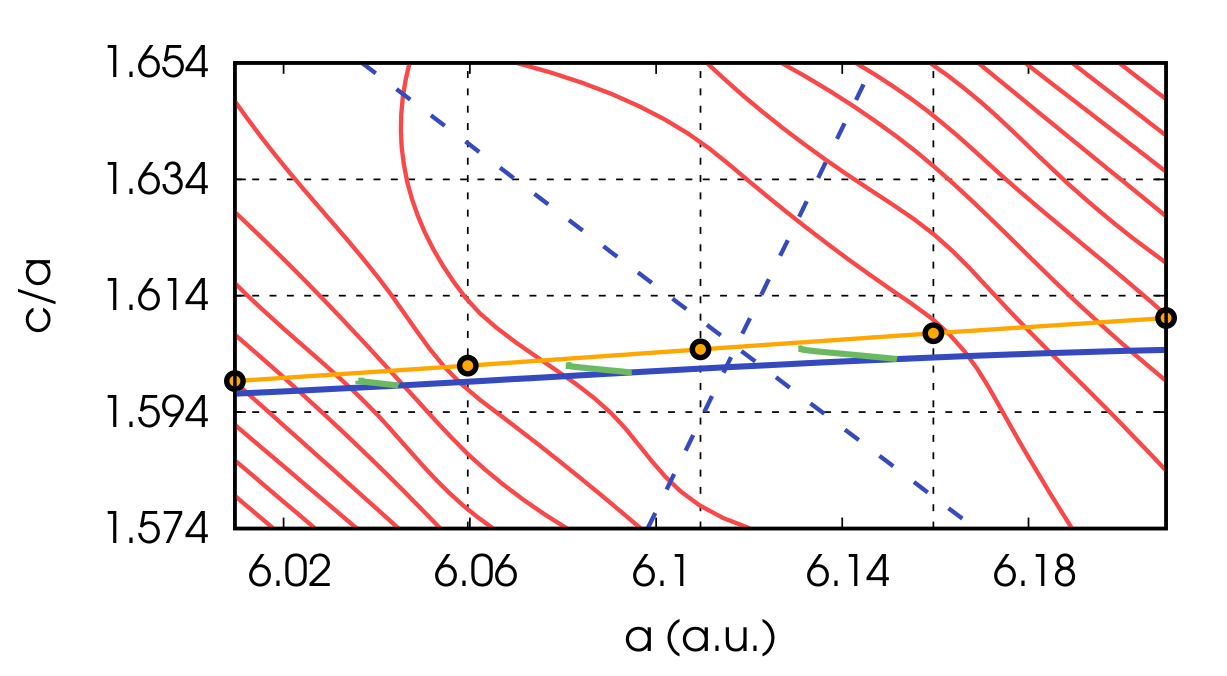}
\caption{Energy contours as a function of the lattice parameter $a$ and the $c/a$ ratio (red lines). The intersection of the dashed blue lines identifies the global energy minimum. The yellow and blue curves represent the configurations where the stress corresponds to a uniform pressure at $0$ K and $700$ K, respectively. The yellow dots indicate the five specific geometries sampled to map the $0$ K stress-pressure curve. Green lines denote the isobars at $0$, $40$, and $80$ kbar. The thin dotted lines show the $5\times 5$ mesh of crystal parameters employed for the Helmholtz free energy calculations.
}
\label{fig:energy_zno}
\end{figure}

\begin{table*}
  \begin{center}
\caption{Equilibrium crystal parameters ($a$, $c/a$, and $u$), bulk modulus ($B_T$), and the first and second pressure derivatives of the bulk modulus ($B_T'$, $B_T^{''}$) for ZnO. The values are obtained from a fourth-order Birch-Murnaghan fit of the total energy. We compare our current PBEsol calculations with previous theoretical studies and experimental data. Results labeled "$0$ + ZPME" indicate crystal parameters obtained by including the zero-point motion energy contribution.
}%
\begin{tabular}{@{}lcccccccc@{}}
\toprule
 &xc& T  &$a$  & $c/a$ & $u$ & $B_T$  & $B_T^{\prime}$ & $B_T^{\prime \prime}$  \\
     & &  (K) & (a.u.) & & & (kbar) &  & (kbar$^{-1}$) \\
\hline
This study (Dojo PP) &PBEsol & $0$ & $6.117$ & $1.605$ & $0.3805$ & $1485$ & $4.08$ & $-0.0063$ \\
This study (Dojo PP) &PBEsol & $0$+ZPME & $6.131$ & $1.605$ & $0.3807$  &  &  & \\
This study (Dojo PP) &PBEsol & $300$ & $6.135$ & $1.604$ & $0.3809$ &  &  & \\
Rostami et al. Ref.~\cite{rostami_anisotropic_2025} &PBEsol & $0$ & $6.098$ & $1.613$ &  & & & \\
Masuki et al. Ref.~\cite{masuki_full_2023} &PBEsol & $0$ & $6.115$ & $1.615$ &  & & & \\
Wu et al. Ref.~\cite{wu_systematic_2005} &LDA & 0 &$6.041$ & $1.616$ & $0.380$ &  &  &    \\
Iwanaga et al. Ref.~\cite{iwanaga_anisotropic_2000}&Expt.& 300 & $6.141$ & $1.602$ &  & $1440$ & &   \\
Yoshio et al. Ref.~\cite{yoshio_crystal_2001}&Expt.& 19 & $6.135$ & $1.603$ & $0.3812$  & & &   \\
Yoshio et al. Ref.~\cite{yoshio_crystal_2001}&Expt.& 293 & $6.140$ & $1.602$ & $0.3815$  & & &   \\
\botrule
\end{tabular}
\label{table:1}
\end{center}
\end{table*}

\section{\label{sec:level2} Theory}

\subsection{Internal and external coordinates}

A periodic solid is characterized by the sizes and shape of its unit cell, which are defined by a set of external crystal parameters ${\xi_i}$ (where $i=1,\dots,N_{ext}$). For the hexagonal wurtzite lattice, $N_{ext}=2$, corresponding to the lattice constant $a$ and the axial ratio $c/a$. The basis is further defined by $3\times N_{at}$ atomic coordinates. While translational invariance reduces the total degrees of freedom by three, the specific space group symmetry of the crystal (for wurtzite, $P6_3mc$) can further constrain these coordinates to a small set of independent internal parameters.

The wurtzite structure is illustrated in Fig.~\ref{fig:wurtzite_struc}. Its geometry is fully determined by the two external parameters, $a$ and $c/a$, and a single internal parameter $u$. This parameter $u$ is sufficient to specify the positions of all $N_{at}=4$ atoms in the unit cell. Physically, the length of the Zn-O bond aligned along the $c$-axis is given by the product $uc$. Under ideal tetrahedral conditions, the internal parameter takes the value $u=3/8$.

We denote the independent internal coordinates as $u_k$, where $k=1,\dots,N_{int}$. To streamline the theoretical description, we introduce the comprehensive notation $\lambda_A$, where $A=1,\dots,N_A$ and 
$N_A=N_{ext}+N_{int}$. This set $\{\lambda_A\}$ encompasses all external and internal parameters required to uniquely define the crystal structure.

 We define also the atomic displacements and the strain tensor. The position of each atom, $\boldsymbol{\tau}_s(u_k)$, is determined by the set of internal parameters. Relative to an equilibrium reference configuration $\lambda_A^{(0)}=(\xi_i^{(0)},u_k^{(0)})$, any arbitrary configuration $\lambda_A=(\xi_i,u_k)$ is characterized by the atomic displacements ${\bf u}_s(u_k)=\boldsymbol{\tau}_s(u_k) - \boldsymbol{\tau}_s(u^{(0)}_k)$ and the strain tensor $\varepsilon_{\alpha\beta} (\{\xi_i,\xi_i^{(0)}\})$.

In the QHA, the Helmholtz free energy F is a function of these 
$\{\lambda_A\}$
variables and the temperature $T$. It can be decomposed into three distinct physical contributions:
\begin{equation}
F(\{\lambda_A\},T)= U(\{\lambda_A\}) + F_{vib}(\{\lambda_A\},T) + F_{el}(\{\lambda_A\},T),
\label{eq:free_ener}
\end{equation}
where $U(\{\lambda_A\})$ is the static DFT energy, 
$F_{vib}(\{\lambda_A\},T)$ the phonon vibrational free energy, and $F_{el}(\{\lambda_A\},T)$ the electronic excitation term. 
Within the QHA, the vibrational component derives from phonon frequencies ${\omega}_{\eta}({\bf q},\{\lambda_A \})$ that depend on $\{\lambda\}$:
\begin{eqnarray} \label{equ4}
F_{vib}(\{\lambda_A\}, T) & =& \frac{1}{2N} \sum_{\mathbf q \eta} \hbar \omega_{\eta} \left(\mathbf q,
\{\lambda_A\} \right) \nonumber\\
& +& {\frac{1}{N \beta}} \sum_{\mathbf q \eta} \ln \left[1 - \exp \left(- \beta \hbar \omega_{\eta}(\mathbf q, \{\lambda_A\})\right) \right].
\end{eqnarray} 
Here $\hbar$ is the reduced Planck's constant,
$N$ is the number of cells of the solid (equal to the number of ${\bf q}$ points), $\beta={\frac{1}{k_B T}}$ where $k_B$ is Boltzmann's constant, ${\bf q}$ denotes phonon wavevectors, and 
$\eta$ indexes vibrational modes. The electronic excitation term $F_{el}(\{\lambda_A\},T)$ follows from the density of states within the rigid-band approximation~\cite{malica_quasi-harmonic_2021}. It gives a non-negligible contribution only on metals.

In order to optimize a crystal structure at temperature $T$ under an external stress $\boldsymbol{\sigma}$, one can minimize the Gibbs energy (in Voigt notation):
\begin{equation}
G_{\boldsymbol{\sigma}}(\{\lambda_A\},T)=F({\lambda_A},T)+\Omega\sum_{j=1}^6 \sigma_j \varepsilon_j(\{\xi_i\}),
\label{gibbsepsilon}
\end{equation}
where $\Omega$ is the volume of one unit cell.
When the stress is a uniform pressure $-p$ this
expression can be simplified as
\begin{eqnarray}
G_{p}(\{\lambda_A\},T)&=&F(\{\lambda_A\},T)-p \Omega(\{\xi_i\}) \\
&=& U(\{\lambda_A\})+F_{vib}(\{\lambda_A\},T)-p \Omega(\{\xi_i\}). \nonumber 
\label{gibbsv}
\end{eqnarray}
The parameters $\{\lambda_A(p,T)\}$ can be determined from the minimization of 
$G_{p}(\{\lambda_A\},T)$.
Thermal expansions, both internal and external, can be defined as:
\begin{equation}
\alpha_A = {1\over \lambda_A} {d \lambda_A \over d T}.
\end{equation}
In general the minimization of 
$G_{p}(\{\lambda_A\},T)$ can be done by computing it in a grid of values, interpolating with a polynomial of $N_A$ variables (usually of fourth degree) and minimizing it. While this procedure is straightforward, it has a bad scaling with $N_A$ and usually simplified approaches are used. 

The zero static internal stress approximation
(ZSISA) treats the internal degree of freedom as functions of the external one and
finds the function $u_{k}(\{\xi_i\})$ by
minimizing the static energy at zero temperature. In this manner
$G_{p}(\{\xi_i\},T)=
G_{p}(\{\xi_i,u_k(\xi_i)\},T)$ and its minimization gives $\xi_i(p,T)$.
The ZSISA theorem, shows that at lowest
order, that is if $F_{vib}(\{\lambda_A\},T)$ is linear polynomial of ${\{\lambda_A\}}$, the external thermal expansion determined within ZSISA is equal to the exact 
one.~\cite{allan_zero_1996} However $u_k(\xi_i(p,T))$ does not gives the correct temperature dependence to the internal thermal expansion.

In this paper, we present an alternative approach generally called full free energy minimization (FFEM) in which, for each set of external parameters, we calculate the free energy in a grid of the internal parameters values and minimize the free energy at fixed values
of $\xi_i$ to obtain $u_k(\xi_i, T)$. In a further step
$ G_{p}(\{\xi_i\},T)=
G_{p}(\{\xi_i,u_k(\xi_i,T)\},T)$ is
minimized giving $\xi_i(p,T)$ and 
$u_k(\xi_i(p,T),T)$ is obtained by interpolation of the previously determined
$u_k(\xi_i,T)$. Numerically, our approach has the same scaling as the minimization the $N_A$ dimensional polynomial, but requires polynomials of a smaller number of variables and is suited to parallelization on multicore machines.

\subsection{Sound speeds and Christoffel equation}

The possible speeds of sound $v$ for waves propagating in an arbitrary direction described by the unit vector ${\bf N}$ are found by solving the Christoffel equation. For a piezoelectric material, this equation is expressed as:~\cite{newnham_properties_2004}
\begin{eqnarray}
\sum_{\gamma}\Bigg[&{1\over \rho}& \sum_{\beta\delta} C^{\bf E}_{\alpha\beta\gamma\delta} {\bf N}_\beta {\bf N}_\delta \\ &+&{1\over \rho}{ \sum_{\lambda\beta}\left( {\bf N}_\lambda e_{\lambda\alpha\beta} {\bf N}_\beta\right) \sum_{\rho\delta} \left({\bf N}_\rho  e_{\rho\gamma\delta} {\bf N}_\delta\right) \over \sum_{\mu\nu} {\bf N}_\mu \epsilon^{(0)}_{\mu\nu}{\bf N}_\nu} \nonumber \\
&-&v^2 \delta_{\alpha\gamma}\Bigg] {\bf u}_\gamma =0, \nonumber
\label{cris_e}
\end{eqnarray}
where $\rho$ is the density,
$C^{\bf E}_{\alpha\beta\gamma\delta}$ are the stress-strain adiabatic elastic constants calculated in zero electric field, 
$e_{\lambda\alpha\beta}$ is the stress piezoelectric tensor, $\epsilon^{(0)}_{\mu\nu}$ is the static dielectric constant and ${\bf u}_\gamma$ is the displacement field that describes the wave.

\subsection{Polarization and Berry phase}
The polarization of a solid, $P_\lambda(\lambda_A)$, is a function of both its internal and external parameters. The difference in polarization between a system characterized by parameters $\lambda_A$ and a reference system with parameters $\lambda^{(0)}_A$ can be determined using the Berry phase approach. This involves calculating the phases $\phi({\bf k},{\bf k}')$ of the determinant of the overlap matrix between the periodic parts of the Bloch wavefunctions, $u_{{\bf k},v}$:
\begin{equation}
S_{v,v'}({\bf k},{\bf k'})=\langle u_{{\bf k},v}|
u_{{\bf k}',v'}\rangle
\end{equation}
as:~\cite{king-smith_theory_1993,resta_macroscopic_1994}
\begin{eqnarray}
P_\lambda(\lambda_A) - 
P_\lambda(\lambda_A^{(0)}) &=& 
- {2 e \over (2\pi)^3}  \Bigg[ \int d^3 k {\partial \over \partial {\bf k}'_\lambda} \phi_{\lambda_A}({\bf k},{\bf k}')\Bigg|_{{\bf k}'={\bf k}} \nonumber \\ &-&
\int d^3 k {\partial \over \partial {\bf k}'_\lambda} \phi_{\lambda_A^{(0)}}({\bf k},{\bf k}')\Bigg|_{{\bf k}'={\bf k}} \Bigg].
\label{polarization_phase}
\end{eqnarray}
\begin{figure}
\centering
\includegraphics[width=\linewidth]{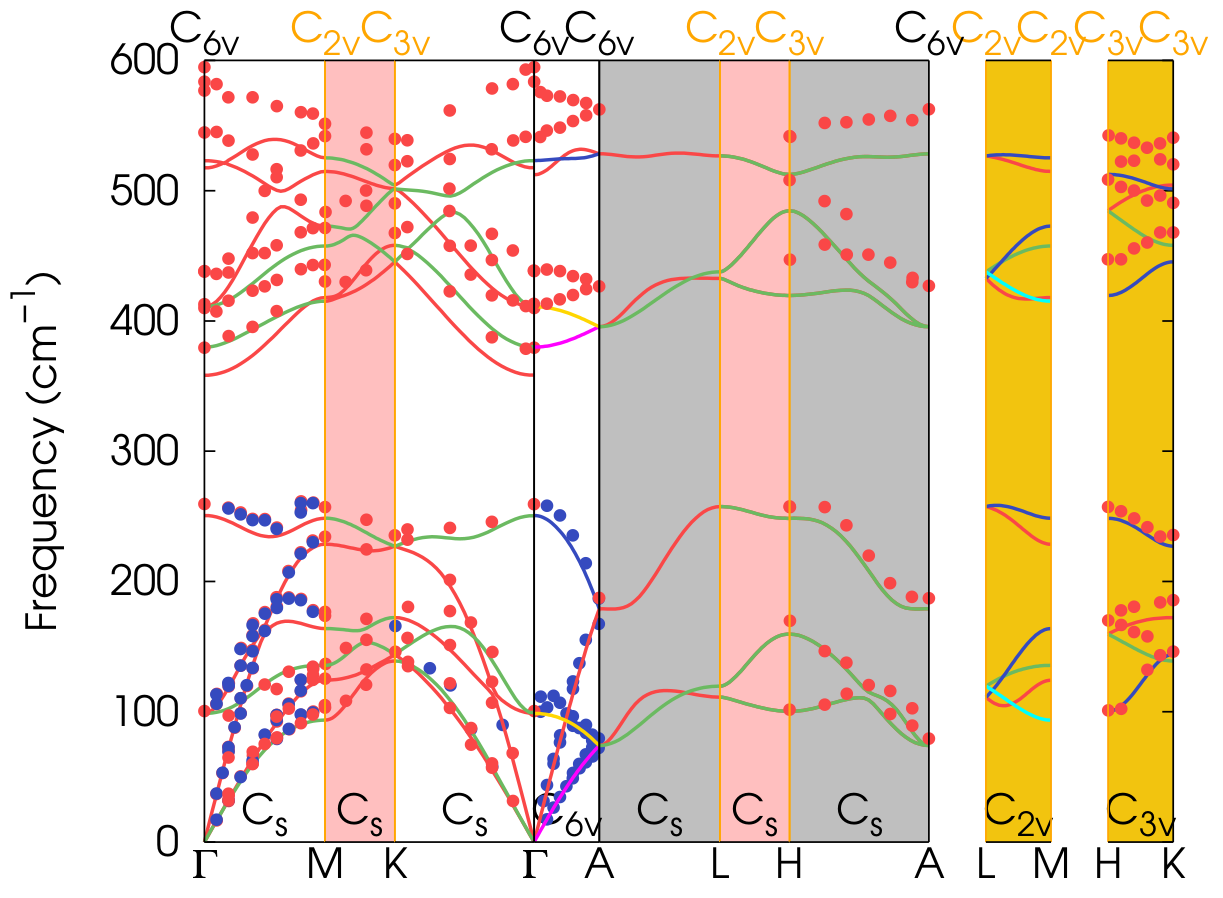}
\caption{PBEsol phonon dispersions of ZnO calculated at the $10$ K equilibrium geometry. The mode symmetries are indicated by color-coding corresponding to the irreducible representations of the point group of the ${\bf q}$ point indicated in the figure. Colored panels represent paths along the Brillouin zone borders; pink and yellow panels utilize projective representations for symmetry classification (refer to the \texttt{thermo\_pw} documentation for detailed color convention definitions). For comparison, experimental inelastic neutron scattering data are included from Ref.~\cite{hewat_lattice_1970} and Ref.~\cite{thoma_lattice_1974} (blue dots), along with data from Ref.~\cite{serrano_phonon_2010} (red dots).}
\label{fig:phon_disp}
\end{figure}

\begin{figure}
\centering
\includegraphics[width=\linewidth]{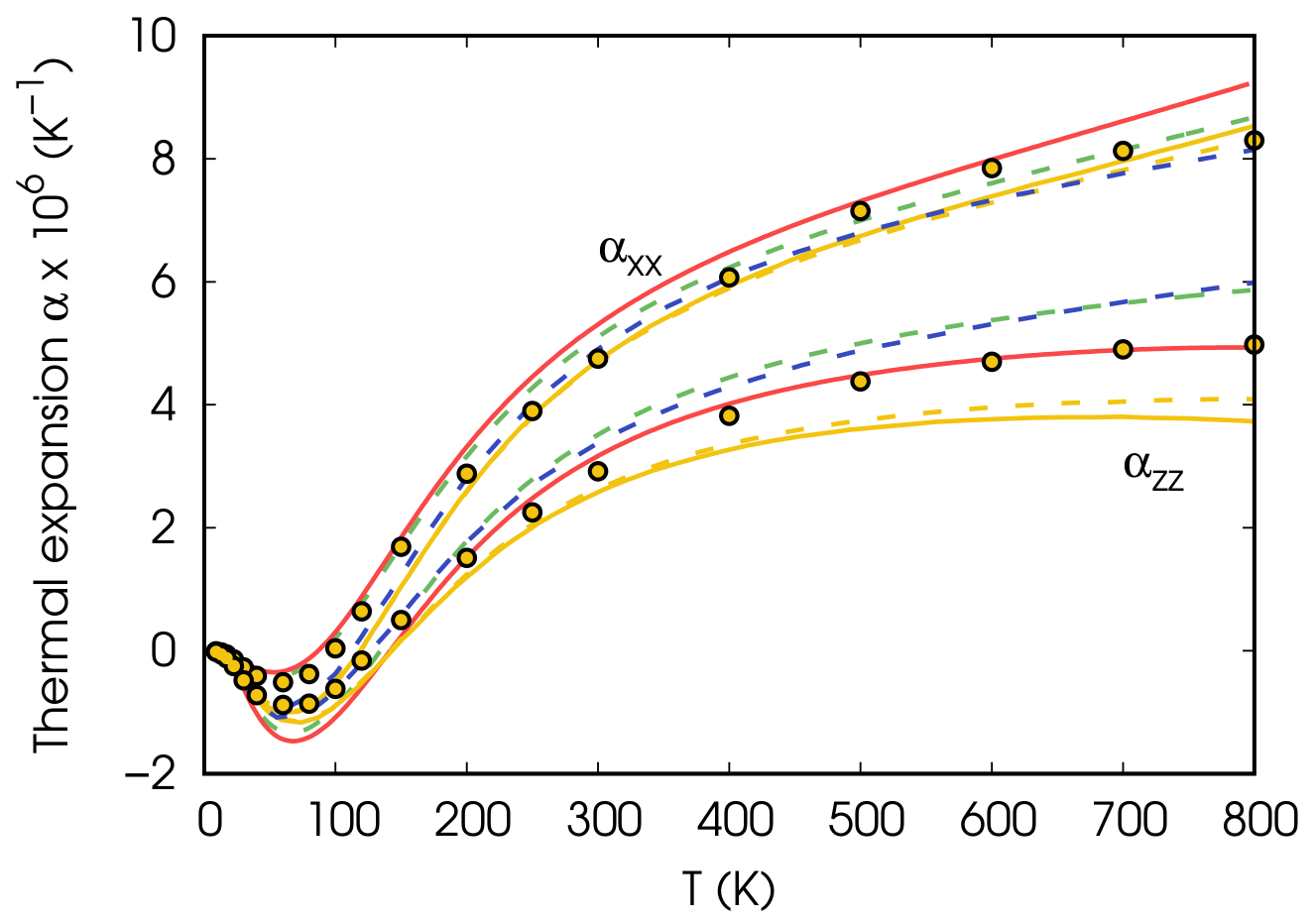}
\caption{PBEsol thermal expansion tensor components, $\alpha_{11}$ and $\alpha_{33}$, of ZnO as a function of temperature. The red and dashed green lines represent our current results obtained using pseudo-dojo pseudopotentials within the FFEM and ZSISA frameworks, respectively. For comparison, we show the ZSISA results from Ref.~\cite{rostami_anisotropic_2025} (dashed blue, also using Dojo PP) and the FFEM (solid yellow) and ZSISA (dashed yellow) results from Ref.~\cite{masuki_full_2023} (calculated with PAW PPs). Experimental data points are taken from Ref.~\cite{ibach_thermal_1969}.}
\label{fig:alpha}
\end{figure}

\subsection{Piezoelectric tensor within FFEM}

The piezoelectric tensor is defined as the derivative of the polarization with respect to the strain $\varepsilon_{\alpha\beta}$ under zero electric field (${\bf E}=0$):
\begin{equation}
e_{\lambda\alpha\beta} = \frac{d P_\lambda (\varepsilon)}{d \varepsilon_{\alpha\beta}}\Bigg|_{{\bf E}=0}
\end{equation}
In this expression, the ions are allowed to relax to their equilibrium positions following the application of strain.

To account for these atomic displacements at finite temperature, we introduce the internal strain parameter $\Gamma_{k\alpha\beta}(T)$ and write the internal degree of freedom
\begin{equation}
u_k(T)  = u_k^{int}(T)+ \sum_{\alpha\beta} \Gamma_{k\alpha\beta}(T) \varepsilon_{\alpha\beta} 
\label{internal_strain}
\end{equation}
we can calculate the total piezoelectric tensor as
\begin{eqnarray}
e_{\lambda\alpha\beta} &=& {\partial {P}_\lambda \over \partial \varepsilon_{\alpha\beta}} +
\sum_k  {\partial {P}_\lambda \over \partial u_k} {\partial u_k \over \partial \varepsilon_{\alpha\beta}} \nonumber \\
&=& \tilde e_{\lambda\alpha\beta} + {e\over \Omega}
\sum_k  Z^*_{k\lambda} {\partial u_k \over \partial \epsilon_{\alpha\beta}} \nonumber \\
&=& \tilde e_{\lambda\alpha\beta} + {e\over \Omega}
\sum_k  Z^*_{k\lambda} \Gamma_{k\alpha\beta}(T)
\label{piezo}
\end{eqnarray}
separating the total piezoelectric response into a clamped-ion term ($\tilde e_{\lambda\alpha\beta}$) and a relaxed-ion term.
The clamped-ion term is calculated by uniformly straining the atomic positions without allowing for further internal relaxation. The second term accounts for the additional polarization induced by atomic relaxations, where $Z^*_{k\lambda}$  are the Born effective charges.
It is important to note that, in general, the applied strains $\varepsilon_{\alpha\beta}$ may lower the symmetry of the crystal, potentially changing the Bravais lattice type. Consequently, the internal degrees of freedom $u_k$ must include all coordinates permitted by the symmetry of the strained structure. For instance, a shear strain in a wurtzite crystal may activate internal displacements $x$ that are otherwise constrained in the unstrained hexagonal phase.

Typically, ${\partial u_k \over \partial \epsilon_{\alpha\beta}}$ are computed only at zero temperature within the ZSISA approximation. However, we consider a more general scheme in which the temperature-dependent coordinates $u_k(T)$ are determined at each temperature by minimizing the Helmholtz free energy for every applied strain 
$\varepsilon_{\alpha\beta}$ within the FFEM framework. In the following sections, we compare the piezoelectric tensor derived using this method with the values obtained via the ZSISA approach.

Born effective charges are defined by the derivative of the polarization with respect to atomic displacements:
\begin{equation}
Z^*_{s\alpha\lambda} = 
{\Omega\over e}{d P_\lambda \over d u_{s\alpha}},
\label{bec}
\end{equation} 
where $e$ is the electron charge and $u_{s\alpha}$ is the displacement of atom $s$ in the direction $\alpha$.
In Eq.~\ref{piezo} we have defined the effective charge associated with the internal parameter $u_k$ as:

\begin{equation}
Z^*_{k\lambda} = {\Omega \over e} \frac{\partial P_\lambda}{\partial u_k} = \sum_{s\alpha} Z^*_{s\alpha\lambda} \frac{d u_{s\alpha}}{d u_k}.
\end{equation}

In a wurtzite material, there are three non-zero piezoelectric coefficients which, in Voigt notation, are denoted as $e_{31}$, $e_{33}$, and $e_{15}$. To calculate these, we consider three specific strain types: two hexagonal strains $(\varepsilon,\varepsilon,0,0,0,0)$ and $(0,0,\varepsilon,0,0,0)$ (under these strains, the cell remains hexagonal, and the internal response is fully described by the parameter $u$ and its strain derivative ${d u \over d \varepsilon}$) and a monoclinic strain: $(0,0,0,0,\varepsilon,0)$. This distortion lowers the Bravais lattice symmetry from hexagonal to monoclinic. While this does not modify $u$ at first order, it allows for a change in the $x$ coordinate of the oxygen atoms.
To determine $e_{15}$, including the relaxed-ion contribution, we require the derivative ${d x \over d \varepsilon}$. 

\subsection{Elastic constants}
Isothermal ECs derive from second derivatives of $F(\varepsilon_{i},T)$:
\begin{equation}
\tilde{C}^T_{ij} = \frac{1}{\Omega} \left.\frac{\partial^2 F}{\partial \varepsilon_{i} \partial \varepsilon_{j}}\right|_T,
\label{tdec}
\end{equation}
evaluated for five distinct strain configurations: $(\varepsilon,0,0,0,0,0)$,
$(0,0,\varepsilon,0,0,0)$, $(\varepsilon,\varepsilon,0,0,0,0)$,
$(\varepsilon,\varepsilon,\varepsilon,0,0,0)$, and $(0,0,0,0,\varepsilon,0)$. In these cases 
${1\over V} {\partial^2 F \over \partial \varepsilon^2}$
gives the following combinations of ECs 
$\tilde C_{11}$, $\tilde C_{33}$, $2\tilde C_{11} +2 \tilde C_{12}$, $2 \tilde C_{11}  + 2 \tilde C_{12} + 4\tilde C_{13} + \tilde C_{33}$, and $\tilde C_{44}$
respectively. 
When the equilibrium reference configuration has a non vanishing stress $\sigma_{ij}$, to obtain the stress-strain ECs we apply the correction~\cite{barron_second-order_1965} (in cartesian notation):
\begin{eqnarray}
C^T_{ijkl} =  \tilde C^T_{ijkl} &-&
{1\over 2} \Big(2 \sigma_{ij} \delta_{kl}
-{1\over 2} \sigma_{ik} \delta_{jl}
-{1\over 2} \sigma_{il} \delta_{jk} \nonumber \\
&-&{1\over 2} \sigma_{jk} \delta_{il}
-{1\over 2} \sigma_{jl} \delta_{ik} \Big),
\label{eqsd2}
\end{eqnarray}
which simplifies for hydrostatic pressure $\sigma_{ij}=-p\delta_{ij}$ to:
\begin{equation}
C^T_{ijkl} = \tilde C^T_{ijkl} + {p \over 2} \left(2 \delta_{i,j} \delta_{k,l}
- \delta_{i,l} \delta_{j,k} - \delta_{i,k} \delta_{j,l}  \right).
\label{eq:correct_p}
\end{equation}

The second derivatives of the free energy are calculated following the methodology described in Ref.~\cite{dal_corso_elastic_2016}. This calculation uses a subset of external parameters $\xi_i$ selected along the $0$~K stress-pressure curve as the reference equilibrium configurations. The specific values of
$\xi_i$
along this curve, along with the corresponding pressure for each configuration, are provided in the Supplemental Material~\cite{supplemental}.

To determine the elastic constants (ECs) for any other parameter set $\xi(p,T)$ at a given temperature $T$ and pressure $p$, we employ a fourth-degree polynomial interpolation. The ECs are obtained by evaluating this polynomial at the specific point on the stress-pressure curve where the volume matches the equilibrium volume $\Omega(\xi(p,T))$.

Adiabatic ECs follow from the transformation:
\begin{equation}
C^S_{ijkl} = C^T_{ijkl} + \frac{TV b_{ij} b_{kl}}{C_V},
\end{equation}
where $b_{ij}$ are the thermal stresses:
\begin{equation}
b_{ij} = - \sum_{kl} C^T_{ijkl} \alpha_{kl},
\end{equation}
where $\alpha_{kl}$ is the thermal expansion tensor.

The QHA elastic constants calculated are consistent within the ZSISA approximations since for each strain the coordinates of the atoms are relaxed using the energy at $0$ K. 

\subsection{Pyroelectric coefficients}

The wurtzite structure lacks an inversion center, which permits a non-zero spontaneous polarization 
$P_\lambda$. This polarization is a function of both the internal and external structural parameters. Given the temperature and pressure dependence of these parameters, denoted as $\lambda_A(p,t)$, we can determine the polarization $P_\lambda(\lambda_A(p,T))$. This yields a functional relationship with pressure and temperature that can be numerically differentiated with respect to temperature to obtain the pyroelectric coefficient:
\begin{equation}
p_\lambda ={d P_\lambda(\lambda_A(p, T)) \over d T}
\end{equation}
This numerical approach allows for the simultaneous inclusion of both primary and secondary contributions, as the parameter set $\lambda_A(p,T)$ accounts for the thermal expansion of the lattice as well as the temperature-driven shifts in internal atomic positions.

In the more common approach, the polarization is expressed via a Taylor expansion:
\begin{equation}
P_\lambda(\lambda_A) = P_{\lambda}(\lambda_A^{(0)}) 
+ {e \over \Omega}
\sum_{s\alpha} Z^*_{s\alpha\lambda} u_{s\alpha}
+ \sum_{\alpha\beta} \tilde e_{\lambda\alpha\beta}
\varepsilon_{\alpha\beta}.
\end{equation}
The pyroelectric coefficient, defined as the temperature derivative of the polarization, is calculated as:
\begin{eqnarray}
p_\lambda(T) = {d P_\lambda(T)\over dT} &=&
{e \over \Omega}\sum_{s\alpha} Z^*_{s\alpha\lambda} {du_{s\alpha}\over d u_k} {d u_k(T) \over dT} \nonumber\\
&+&  \sum_{\alpha\beta} \tilde e_{\lambda\alpha\beta} \alpha_{\alpha\beta}. 
\end{eqnarray}
By decomposing the variation of the internal parameters into clamped-lattice ($u^{int}$) and strain-induced ($u^{ext}$) components according to Eq.~\ref{internal_strain}:
\begin{eqnarray}
{d u_k(T) \over dT}&=& {d u^{int}_k(T) \over dT}
+{d u^{ext}_k(T) \over dT} =
{d u^{int}_k(T) \over dT} \nonumber \\ &+& \sum_{\alpha\beta} \Gamma_{k\alpha\beta}(T) \alpha_{\alpha\beta}
\label{ugamma}
\end{eqnarray}
the pyroelectric tensor can be written as the sum of three distinct contributions:
\begin{eqnarray}
p_\lambda(T) = {d P_\lambda(T)\over dT} &=&
{e\over \Omega} \sum_{k} Z^*_{k\lambda} {d u^{int}_k(T) \over dT} \nonumber\\
&+&
{e\over \Omega} \sum_{k} Z^*_{k\lambda} {d u^{ext}_k(T) \over dT} \nonumber\\
&+&  \sum_{\alpha\beta} \tilde e_{\lambda\alpha\beta} \alpha_{\alpha\beta}, 
\label{pyro}
\end{eqnarray}
Using the relationship in Eq.~\ref{ugamma} and the definition of the total piezoelectric tensor (Eq.~\ref{piezo}), this simplifies to:
\begin{eqnarray}
p_\lambda(T) &=&
{e\over \Omega} \sum_{k} Z^*_{k\lambda} {d u^{int}_k(T) \over dT} \nonumber\\
&+&  \sum_{\alpha\beta}  e_{\lambda\alpha\beta} \alpha_{\alpha\beta}, 
\label{pyro1}
\end{eqnarray}
In this formulation, we neglect the explicit temperature dependence of the Born effective charges, the clamped-ions piezoelectric tensor, and the internal strain parameters; the piezoelectric tensor used is the total one.
Experimentally, the pyroelectric tensor is typically divided into primary and secondary contributions. The primary contribution corresponds to the first term in Eq.~\ref{pyro1} (the clamped-lattice effect), while the secondary contribution arises from the lattice deformation, represented by the second term.

\begin{figure}
\centering
\includegraphics[width=\linewidth]{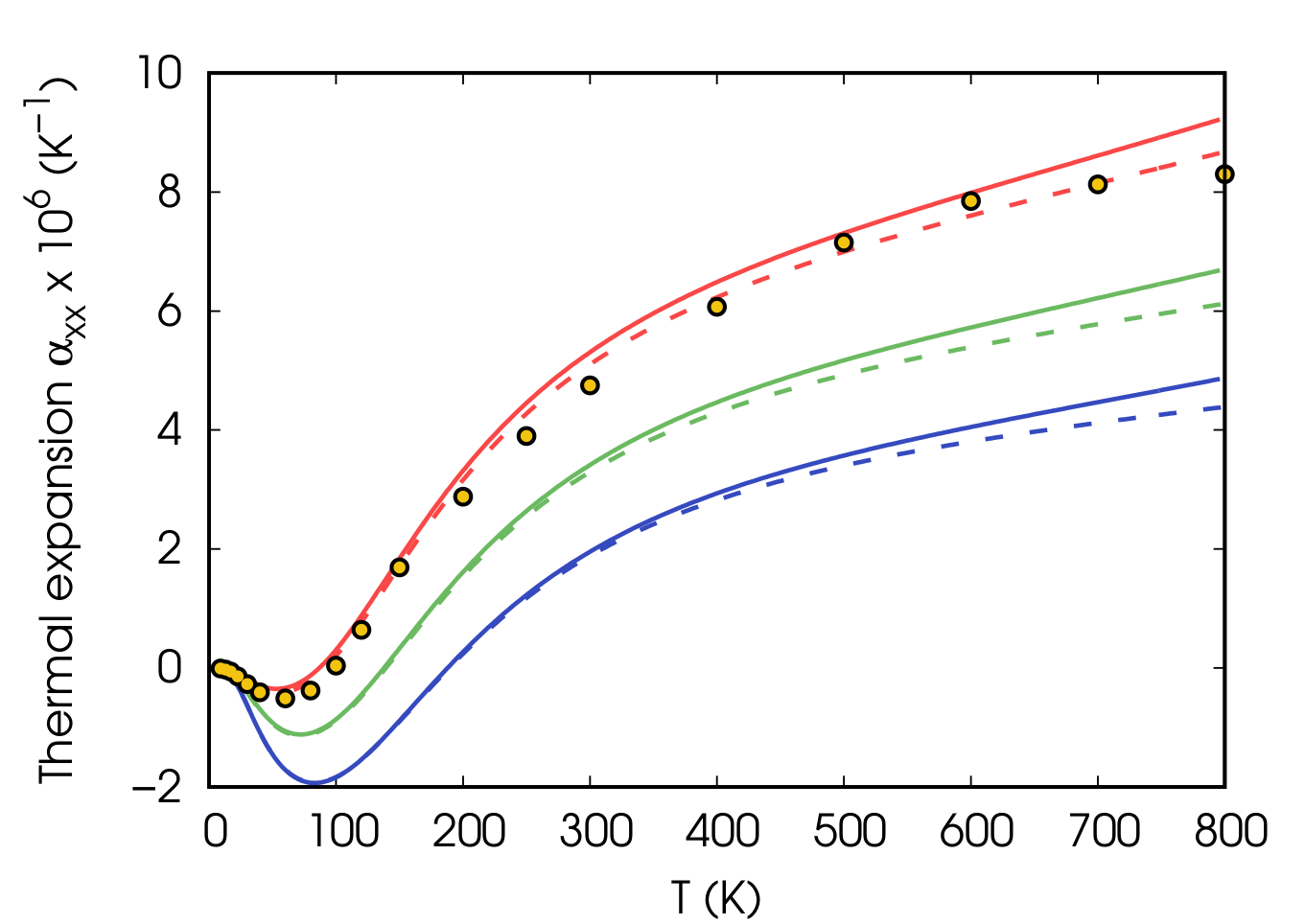}
\caption{PBEsol thermal expansion coefficient $\alpha_{xx}$ of ZnO as a function of temperature, calculated at pressures of $0$ kbar (red), $40$ kbar (green), and $80$ kbar (blue). Continuous lines represent the results obtained using the FFEM, while dashed lines correspond to the ZSISA approximation. The comparison illustrates the suppression of thermal expansion with increasing pressure and the divergence between the two methodological frameworks as temperature rises.}
\label{fig:alpha_xx}
\end{figure}

\begin{figure}
\centering
\includegraphics[width=\linewidth]{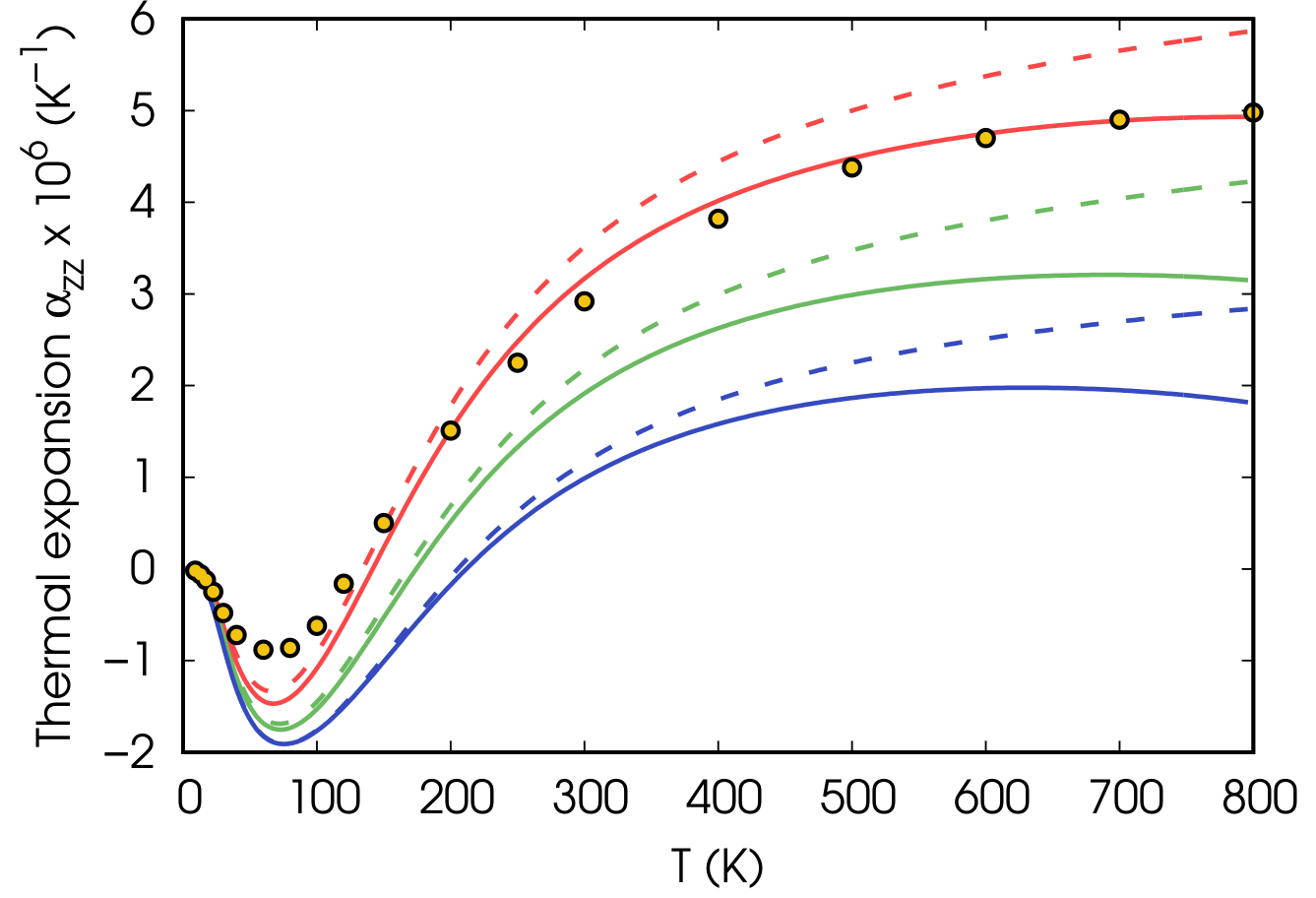}
\caption{PBEsol thermal expansion coefficient $\alpha_{zz}$ of ZnO as a function of temperature, calculated at pressures of $0$ kbar (red), $40$ kbar (green), and $80$ kbar (blue). Continuous lines denote the FFEM results, whereas dashed lines represent the ZSISA approximation.}
\label{fig:alpha_zz}
\end{figure}

For the specific case of the wurtzite structure, polarization is restricted to the $c$-axis and there is only one internal degree of freedom ($k=1$). Since 
${d u_{s3} \over d u} = c$
for two atoms (either the anions or the cations) and zero for the others, we find $Z^*_{13}=2 c Z^*_{O33}$ (or equivalently
$Z^*_{13}= 2 c Z^*_{Zn 33}$).
Eq.~\ref{pyro1} becomes:
\begin{equation}
p_3 = {e\over \Omega} Z^*_{13} {d u^{int}(T) \over dT} + 2 e_{31} \alpha_{11} + e_{33} \alpha_{33}.
\label{pyro1_wurt}
\end{equation}

\begin{table*}
  \caption{Elastic constants $C_{ij}$ of ZnO compared with experimental results and previous theoretical calculations. The adiabatic QHA values reported at $4$ K include the zero-point motion energy (ZPME) contribution. The influence of thermal effects is highlighted by the inclusion of the $300$ K elastic constants. For comparison, frozen-ion elastic constants are reported, obtained by applying a uniform strain to the atomic positions while omitting internal atomic relaxation and maintaining the $0$ K equilibrium geometry. In the references, only the first author is cited for brevity.}
\begin{ruledtabular}
\begin{tabular}{cccccccccc}
 & T &$a$ & $\frac{c}{a}$ & $u$ & $C_{11}$  & $C_{12}$ & $C_{13}$ & $C_{33}$ & $C_{44}$  \\
  & (K) & (a.u.) &  & & (kbar) & (kbar) & (kbar) & (kbar) & (kbar) \\
\hline
 This study (PBEsol) & $0$ & $6.117$ & $1.605$ & $0.3805$& $2152$ & $1274$ & $1111$ & $2100$ & $405$   \\
Clamped ions (PBEsol) (QSA) & $4$ & $6.131$ & $1.604$ & $0.3807$ & $2752$ & $950$ & $702$ & $2801$ & $574$  \\
This study (PBEsol) {(${\bf E}=0$)}& $4$ & $6.131$ & $1.604$ & $0.3807$& $2096$ & $1242$ & $1083$ & $2051$ & $396$  \\
This study (PBEsol) {(${\bf D}=0$)}& $4$ & $6.131$ & $1.604$ & $0.3807$& $2133$ & $1279$ & $1008$ & $2201$ & $419$  \\
This study (PBEsol) {(${\bf E}=0$)}& $300$ & $6.135$ & $1.604$ & $0.3807$& $2014$ & $1189$ & $1037$ & $1993$ & $348$  \\
 Rostami Ref.~\cite{rostami_anisotropic_2025} (PBEsol) & $1$ & $6.098$ & $1.613$ & & $1996$ & $1218$ & $1068$ & $2114$ & $352$  \\
  Rostami Ref.~\cite{rostami_anisotropic_2025} (PBEsol) & $300$ &  & & & $1934$ & $1182$ & $1026$ & $2059$ & $352$  \\
 Wang Ref.~\cite{wang_thermodynamic_2014} (LDA) & $0$ & $6.016$ & $1.617$ &  & $2235.1$ & $1398.8$ & $1499.6$ & $2921.8$ & $367.2$   \\
 Wang Ref.~\cite{wang_thermodynamic_2014} (PBEsol) & $0$ & $6.118$ & $1.614$ &  & $2010.0$ & $1244.9$ & $1195.7$ & $2353.8$ & $347.3$   \\
 Wang Ref.~\cite{wang_thermodynamic_2014} (PBE) & $0$ & $6.202$ & $1.613$ &  & $1911.6$ & $1119.6$ & $856.3$ & $1905.7$ & $368.9$  \\
  Lakel Ref.~\cite{lakel_pressure_unpublished}(PBE) & $0$ & $6.200$ & $1.614$ & & $1889.6$ & $1103.2$ & $904.2$ & $2009.7$ & $368.7$ \\
  Shein Ref.~\cite{shein_elastic_2007} (PBE) & $0$ & $6.181$ & $1.611$ &  & $1954$ & $1112$ & $925$ &
   $1998$ & $396$ \\
Wu Ref.~\cite{wu_systematic_2005} (LDA) & $0$ & $6.041$ & $1.616$ & $0.380$ &$2260$ & $1390$ & $1230$ & $2420$ & $400$  \\
Clamped ions Wu Ref.~\cite{wu_systematic_2005} (LDA) & $0$ & $6.041$ & $1.616$ & $0.380$ &$3050$ & $1070$ & $770$ & $3300$ & $620$  \\
Liu Ref.~\cite{liu_internal_2018} (LDA) & $0$ & $6.041$ & $1.616$ & $0.379$ &  & & $1220$ & $2260$  \\
Bateman Ref.~\cite{bateman_elastic_1962} (Expt.) (${\bf E}=0$) & $298$ & & & & $2097$ & $1211$ & $1051$ & $2109$ & $425$  \\
Kobiakov Ref.~\cite{kobiakov_elastic_1980} (Expt.) (${\bf E}=0$) & $300$ & & & & $2070$ & $1177$ & $1061$ & $2095$ & $448$  \\
Kobiakov Ref.~\cite{kobiakov_elastic_1980}(Expt.) (${\bf D}=0$) & $300$ &  &  & & $2096$ & $1204$ & $1013$ & $2210$ & $461$  \\
Azuhata Ref.~\cite{azuhata_brillouin_2003} (Expt.) & $0$ &  &  &  & $1900$ & $1100$ & $900$ & $1960$ & $390$  \\
\end{tabular}
\label{table:2}
\end{ruledtabular}
\label{tab:elastic}
\end{table*}

\subsection{Dielectric constants}

The static dielectric constant $\epsilon^{(0)}_{\alpha\beta}$ of an insulator can be expressed by combining the electronic (clamped-ion) contribution with the ionic lattice response. The electronic part, $\epsilon^{\infty}_{\alpha\beta}$, is measured at frequencies high enough to surpass phonon resonances but low enough to remain below electronic excitations.

By utilizing the eigenvalues  ($ \omega^2_{{\bf 0},\eta}$) and the mass-normalized eigenvectors of the dynamical matrix (${\bf u}^\eta_{s\gamma}({\bf 0})$) at the $\Gamma$ point of the Brillouin zone, the static dielectric constant is given by:~\cite{gonze_dynamical_1997}
\begin{equation}
\epsilon_{\alpha\beta}=\epsilon^\infty_{\alpha\beta} + {4 \pi \over \Omega} \sum_\eta {\sum_{s\gamma} Z^*_{s\gamma\alpha} {\bf u}^\eta_{s\gamma}({\bf 0}) \sum_{s'\delta} Z^*_{s'\delta\beta} {\bf u}^\eta_{s'\delta}({\bf 0}) \over \omega^2_{{\bf 0},\eta}},
\label{epsilon0}
\end{equation}
where the sum excludes the three acoustic modes.
$\epsilon^\infty_{\alpha\beta}$
and the Born effective charges 
$Z^*_{s'\delta\beta}$ are calculated by DFPT from the derivatives of the polarization:~\cite{rmp}
\begin{equation}
\epsilon^\infty_{\alpha \beta}=\delta_{\alpha \beta}+4 \pi \frac{d P_\beta}{d E_\alpha},
\end{equation}
and Eq.~\ref{bec}.


\subsection{Elastic constant at constant electric displacement}

The elastic constants at constant (vanishing) displacement field (${\bf D}$) are defined as:\cite{wu_systematic_2005}
\begin{equation}
C^{{\bf D}}_{\alpha\beta\gamma\delta} = 
C^{{\bf E}}_{\alpha\beta\gamma\delta} + \sum_{\lambda\rho} e_{\lambda\alpha\beta}  \epsilon_{\lambda\rho}^{-1} e_{\rho\gamma\delta}.
\end{equation}
In the case of a wurtzite structure the corrections
become:
\begin{eqnarray}
C^{\bf D}_{11} &=& C^{\bf E}_{11} + {(e_{31})^2 \over \epsilon^{(0)}_{3}}, \nonumber \\
C^{\bf D}_{12} &=& C^{\bf E}_{12} + {(e_{31})^2 \over \epsilon^{(0)}_{3}}, \nonumber  \\
C^{\bf D}_{13} &=& C^{\bf E}_{13} + {e_{31} e_{33} \over \epsilon^{(0)}_{3}}, \nonumber  \\
C^{\bf D}_{33} &=& C^{\bf E}_{33} + {(e_{33})^2 \over \epsilon^{(0)}_{3}}, \nonumber  \\
C^{\bf D}_{44} &=& C^{\bf E}_{44} + {(e_{15})^2 \over \epsilon^{(0)}_{1}}.
\label{c_d_e}
\end{eqnarray}

\section{\label{sec:level3}Computational details}

First-principles calculations were performed within DFT and DFPT using the Quantum ESPRESSO distribution~\cite{qe1, qe2}. We employed the PBEsol exchange-correlation functional~\cite{pbesol} throughout. To maintain consistency with recent literature, electron-ion interactions were described by PseudoDojo norm-conserving pseudopotentials~\cite{van_setten_pseudodojo_2018}. For zinc, the pseudopotential explicitly includes $3s$ and $3p$ semicore states alongside the $3d$ and $4s$ valence electrons ($20$ valence electrons per atom); for oxygen, $2s$ and $2p$ states were included in the valence configuration.

Wavefunctions and charge density were expanded in plane waves with kinetic energy cutoffs of $100$ Ry and $400$ Ry, respectively. Brillouin zone integration was performed using a centered $8\times 8\times 6$ {\bf k}-point mesh. All calculations were executed on the Leonardo supercomputer (CINECA) utilizing a GPU-accelerated version of the \texttt{thermo\_pw} package~\cite{gongAlternativeGPUAcceleration2025}.

Equilibrium lattice parameters were determined by sampling the total energy on a $5\times 5$ grid of     $(a,c/a)$ values, spanning a pressure range from $-64$ kbar to $94$ kbar. The grid was centered at $a=6.1095$ a.u. and $c/a=1.614$, with step sizes of $0.05$ a.u. and $0.02$, respectively. Within the ZSISA, the internal coordinate $u$ was optimized at each grid point by minimizing the total energy.

Phonon dispersions and Helmholtz (or Gibbs) free energies were computed at each grid point, interpolated via a fourth-order polynomial, and minimized at each temperature and pressure. This procedure allowed us to define the stress-pressure curves at various temperatures. To determine the bulk modulus and its pressure derivatives, we selected five points along the $0$ K stress-pressure curve (see Supplemental Material~\cite{supplemental}) and fitted the results to a fourth-order Birch-Murnaghan equation of state.~\cite{birch_finite_1947}
\begin{figure}
\centering
\includegraphics[width=\linewidth]{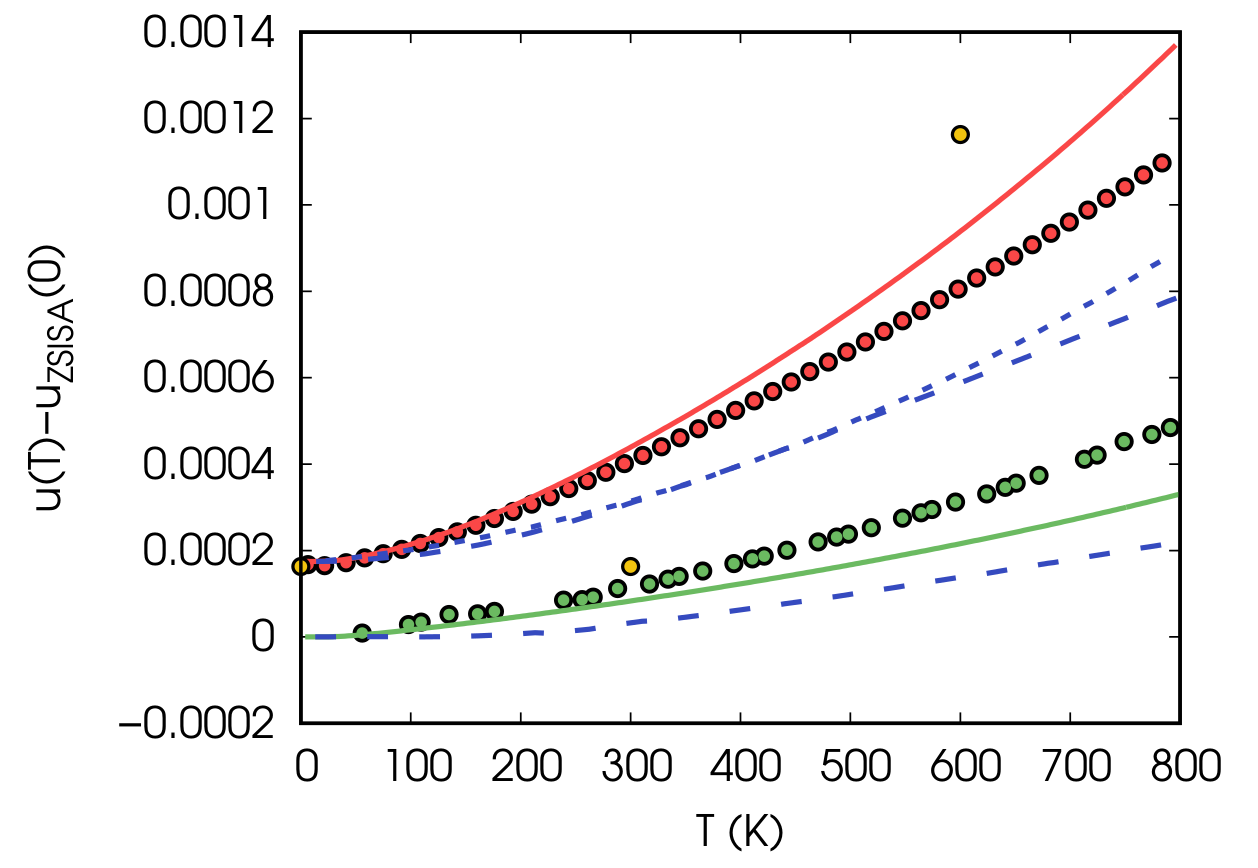}
\caption{PBEsol internal parameter $u(T)$ of ZnO as a function of temperature. Results are shown for the ZSISA (green line) and FFEM (red line) frameworks. To facilitate comparison, the vertical scale is shifted such that $u_{ZSISA}(0)=0$. We include data from Ref.~\cite{liu_erratum_2018}, where the red dots represent the total variation ($u_{int}+u_{ext}$) and green dots represent the strain-induced part ($u_{ext}$). Blue dashed lines (from bottom to top) indicate the ZSISA, Grüneisen, and refined Grüneisen ($E^{(3)}+F_{vib}^{(2)}$) results from Ref.~\cite{liu_internal_2018}. For consistency, all datasets are aligned such that $u_{ZSISA}(0)=0$, while the $u(0)$ values are shifted to match our FFEM results. Experimental data from Ref.~\cite{albertsson_atomic_1989} (yellow dots) are similarly aligned to our FFEM u(0) value.}
\label{fig:ut_zno}
\end{figure}

\begin{figure}
\centering
\includegraphics[width=\linewidth]{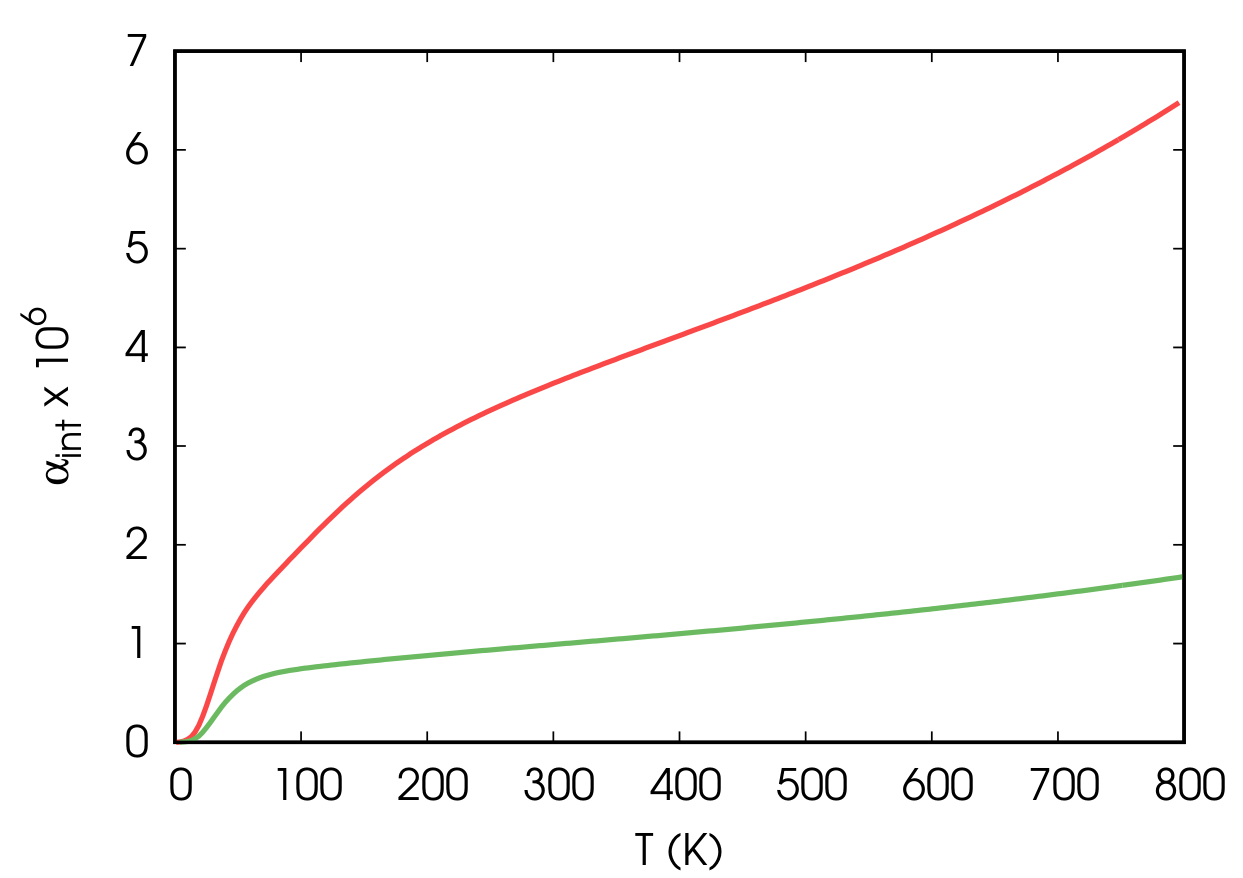}
\caption{PBEsol internal thermal expansion of ZnO as a function of temperature, defined as the temperature derivative of the internal parameter (${{1\over u}{d u \over d T}}$). The curves are obtained by numerical differentiation of the FFEM (red) and ZSISA (green) data presented in Fig.~\ref{fig:ut_zno}.}
\label{fig:alpha_int_zno}
\end{figure}

Temperature-dependent elastic constants (TDECs) were evaluated using second derivatives of the free energy with respect to strain. For each of the five selected equilibrium geometries, five distinct strain types were applied: base-centered orthorhombic $(\varepsilon,0,0,0,0,0)$, hexagonal $(1)$ $(0,0,\varepsilon,0,0,0)$, $(2)$ $(\varepsilon,\varepsilon,0,0,0,0)$, and $(3)$ 
$(\varepsilon,\varepsilon,\varepsilon,0,0,0)$, monoclinic $(0,0,0,0,\varepsilon,0)$. Each strain was sampled at six values ($\varepsilon \in  [-0.0125,0.0125]$ with 
$\Delta\varepsilon=0.005$), requiring phonon
calculations for $30$ distorted configurations per equilibrium geometry ($150$ in total).

The piezoelectric tensor was determined by calculating the Berry phase polarization for the hexagonal $(1)$ and $(2)$ and monoclinic strains. For these, we sampled four strain values ($\varepsilon \in [-0.0075,0.0075]$, $\Delta\varepsilon=0.005$). We performed these calculations both on the full $5 \times 5$ grid 
($1.500$ Berry phase calculations) and the five points along the stress-pressure curve ($60$ calculations).
The Berry phase calculations were performed with the following discretization: The one-dimensional {\bf k}-point lines (along the direction of the polarization) were discretized using $21$ points, while the directions perpendicular to the integration string, we used the same ${\bf k}$-point mesh density employed during the self-consistent cycles.

For the FFEM approach, the temperature-dependent external and internal parameters were determined by sampling five values of $u$ for each $(a,c/a)$ pair, totaling $125$ phonon dispersion calculations for the primary grid.

\begin{table*}
  \begin{center}
\caption{
Clamped-ion and relaxed-ion piezoelectric tensor components ($e_{31}$, $e_{33}$, and $e_{15}$) of ZnO, reported in units of C/m$^2$. The results from the present work are compared with available literature values.}
\begin{tabular}{lccccccc}
\toprule
 &  T (K)  &$\tilde e_{31}$  & $\tilde e_{33}$ & $\tilde e_{15}$ & $e_{31}$  & $e_{33}$  & $e_{15}$  \\
 \hline
 This work (PBEsol)  &  $4$  &$0.36$  & $-0.73$ & $0.38$ & $-0.62$  & $1.25$  & $-0.48$  \\
  This work (PBEsol)  &  $300$  & $0.36$ & $-0.73$ & $0.38$ & $-0.61$  & $1.23$  & $-0.47$  \\
Ref.~\cite{wu_systematic_2005} (LDA)  &  $0$  &$0.37$  & $-0.75$ & $0.39$ & $-0.67$  & $1.28$  & $-0.53$  \\
Ref.~\cite{catti_full_2003} (HF)  &  $0$  &  &  &  & $-0.55$  & $1.19$  & $-0.46$  \\
Ref.~\cite{noel_polarization_2001} (HF)  &  $0$  & $0.22$ & $-0.44$ &  & $-0.53$  & $1.19$  & \\
Ref.~\cite{dal_corso_ab_1994} (LDA)  &  $0$  &$0.37$  & $-0.58$ & & $-0.51$  & $1.21$  &  \\
Ref.~\cite{kobiakov_elastic_1980} (Expt.) &  $0$  &  &  & & $-0.62$  & $0.96$  & $-0.37$  \\
\hline
\botrule
\end{tabular}
\label{table:3}
\end{center}
\end{table*}

To determine $u(T)$ for each strained configuration in the piezoelectric calculations, we computed the free energy for five values of the internal parameter 
($u$ or $x$) for each of the $12$ distortions. This necessitated an additional $60$ phonon calculations per equilibrium geometry, repeated for the three central geometries along the stress-pressure curve. Finally, all phonon frequencies were obtained via DFPT on a $4\times 4\times 4$ {\bf q}-point mesh and Fourier-interpolated onto a dense 
$200\times200\times 200$ mesh for thermodynamic integration. This results in $21$, $12$, and $24$ symmetry inequivalent ${\bf q}$ points for the orthorhombic, hexagonal and monoclinic cells, respectively. 

\section{\label{sec:level4}Results and discussion}

We begin by summarizing the results obtained at zero temperature. Table~\ref{table:1} reports the equilibrium lattice parameters ($a$ and $c/a$), the internal coordinate ($u$), the bulk modulus, and its first and second pressure derivatives. 
Our results are in good agreement with existing literature. The calculated $a$ parameter deviates by only $0.3\%$ from experimental values—a discrepancy that further diminishes when accounting for zero-point motion effects (ZPME). While our $a$ parameter is consistent with that of Ref.~\cite{masuki_full_2023}, it is slightly larger than the value reported in Ref.~\cite{rostami_anisotropic_2025}. Conversely, our $c/a$ ratio is slightly lower than those found in both references, placing it in closer alignment with experimental data.

Figure~\ref{fig:energy_zno} illustrates the energy contour levels at zero temperature, alongside the stress-pressure curves--defined as the paths where the stress is a uniform pressure--at $0$ K and $700$ K. We also plot the isobars connecting these two curves at pressures of $0$, $40$, and $80$ kbar. The energy minimum corresponds to the intersection of the two dashed lines. Notably, the $0$ kbar isobar originates from the free energy minimum at the lowest calculated temperature ($7$ K); this point incorporates ZPME and is consequently slightly shifted from the static energy minimum. The points indicated in the figure represent the geometries used to calculate the quasiharmonic (QHA) elastic constants and the piezoelectric tensor within the ZSISA approximation. For the three central points, we additionally calculated the piezoelectric tensor using the FFEM approach to determine the strain induced change of the internal parameter $u$ (or $x$).

\begin{figure}
\centering
\includegraphics[width=\linewidth]{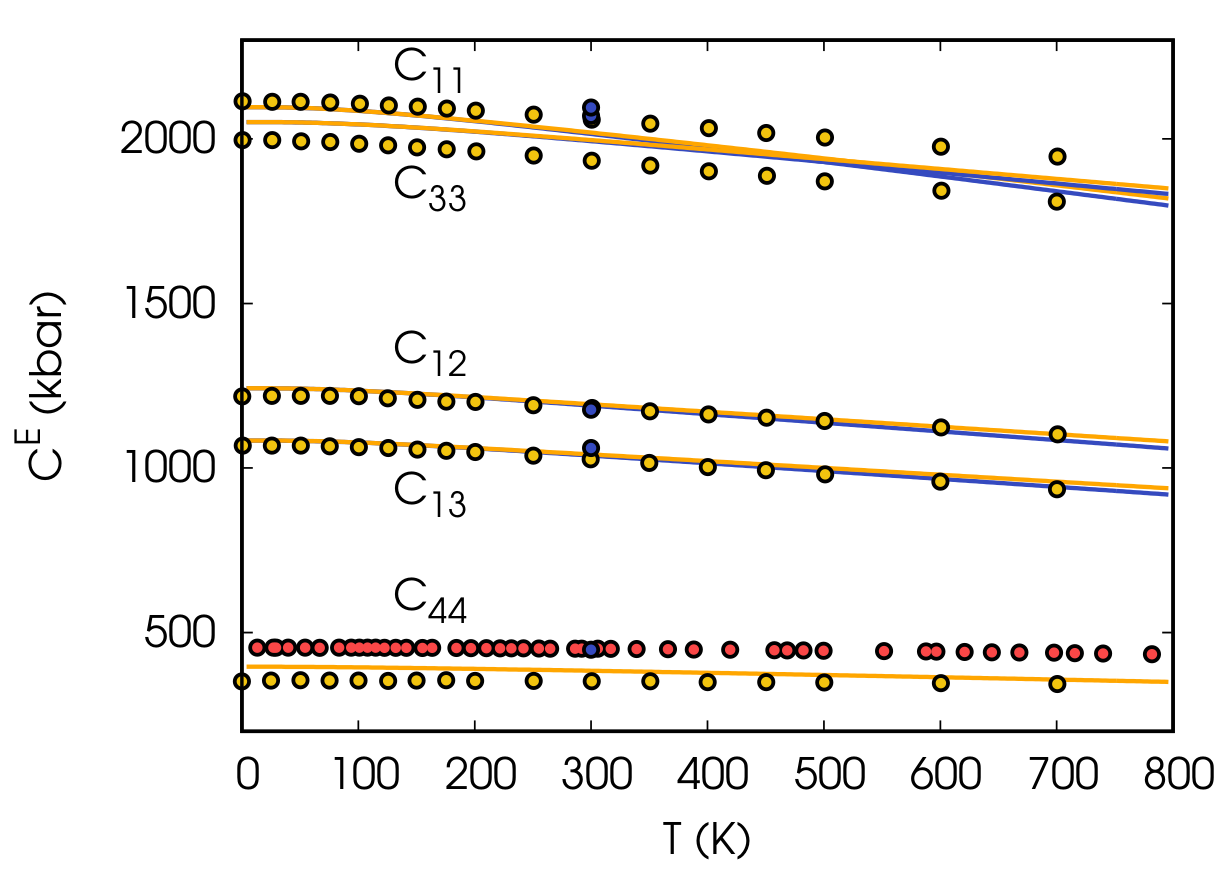}
\caption{PBEsol elastic constants of ZnO (calculated under constant electric field, ${\bf E}$, conditions) as a function of temperature at $0$ kbar. Orange and blue lines represent the adiabatic ($C^S_{ij}$) and isothermal ($C^T_{ij}$) elastic constants, respectively, obtained within the V-ZSISA framework by interpolating results across five geometries along the stress-pressure curve. Yellow dots indicate the PBEsol results from Ref.~\cite{rostami_anisotropic_2025}
(with $C_{33}>C_{11}$. Experimental data at constant ${\bf E}$ are provided from Ref.~\cite{kobiakov_elastic_1980} (red dots), with specific $300$ K experimental values from the same source are highlighted as blue dots.}
\label{fig:el_cons_qha}
\end{figure}

\begin{figure}
\centering
\includegraphics[width=\linewidth]{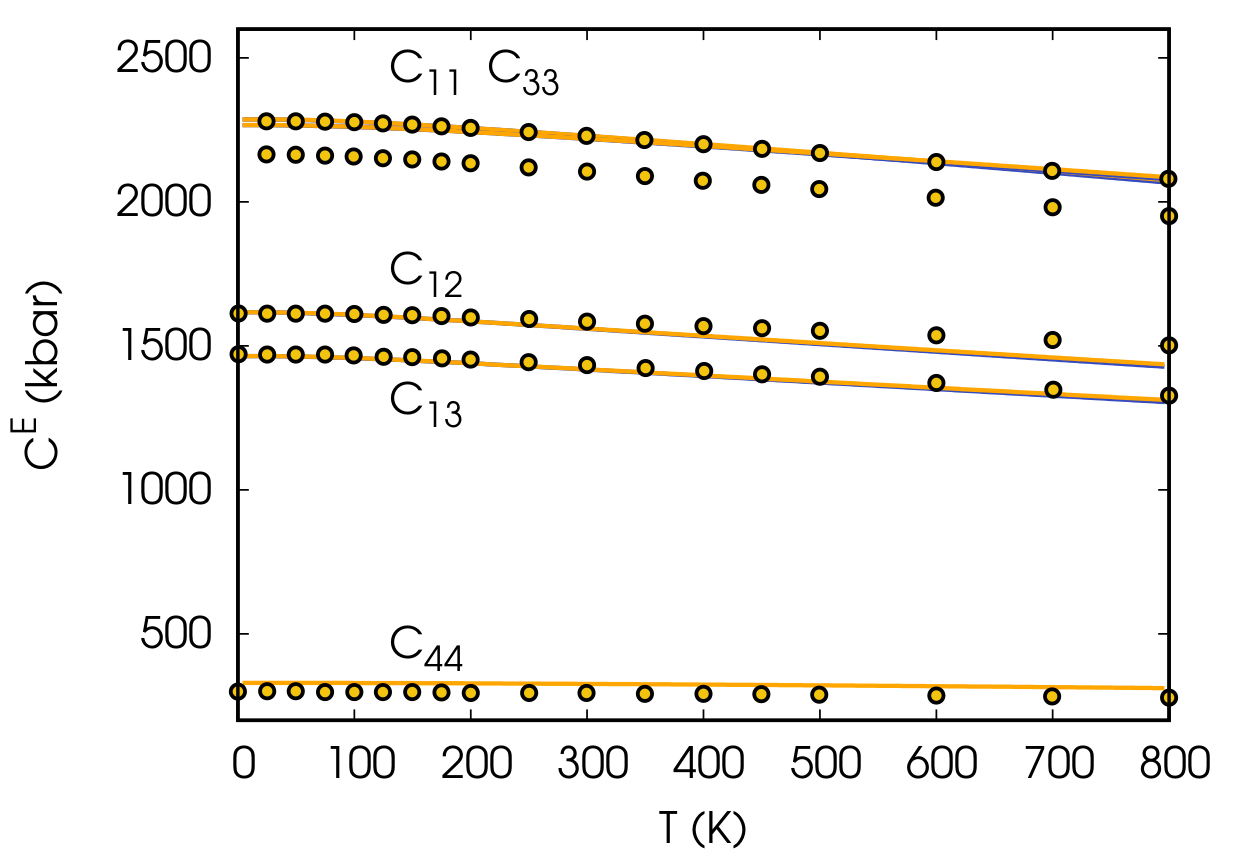}
\caption{PBEsol elastic constants of ZnO as a function of temperature under a hydrostatic pressure of $80$ kbar, calculated at constant electric field (${\bf E}$). Orange and blue lines represent the adiabatic 
($C^S_{ij}$) and isothermal ($C^T_{ij}$) constants, respectively, determined within the V-ZSISA framework using interpolation across five reference geometries. Yellow dots indicate the PBEsol calculations from Ref.~\cite{rostami_anisotropic_2025}. The results from Ref.~\cite{rostami_anisotropic_2025} show $C_{33}>C_{11}$ also in this high-pressure regime.}
\label{fig:el_cons_qha_80}
\end{figure}

The PBEsol phonon dispersions, calculated using the current pseudopotential, are shown in Fig.~\ref{fig:phon_disp}. These dispersions are compared with inelastic neutron scattering data from Ref.~\cite{hewat_lattice_1970} and Ref.~\cite{thoma_lattice_1974} (measured at $300$ K) and the data Ref.~\cite{serrano_phonon_2010} (measured at $10$ K). The displayed dispersions represent frequencies interpolated from DFPT calculations on a $5\times 5$ grid, using crystal parameters corresponding to $10$ K.

The PBEsol results are slightly lower than experimental values, particularly across the optical branches. While the LDA phonon calculations in Ref.~\cite{serrano_pressure_2004} show better agreement with experiment, this discrepancy may stem from both the choice of exchange-correlation functional and the specific pseudopotentials employed. Indeed, norm-conserving pseudopotentials have been shown to exhibit larger variations relative to all-electron calculations compared to ultrasoft or PAW pseudopotentials.

Figure~\ref{fig:alpha} displays the two components of the thermal expansion tensor, $\alpha_{xx}$ and $\alpha_{zz}$, calculated on a $5\times 5$ mesh using both the ZSISA and FFEM frameworks. In ZnO, $\alpha_{xx}$ is greater than $\alpha_{zz}$, which corresponds to a decrease in the $c/a$ ratio as temperature increases. Our PBEsol-based results show strong agreement with experimental data, particularly for $\alpha_{xx}$. While the ZSISA results for 
$\alpha_{zz}$
slightly overestimate experimental values, the FFEM results are in excellent agreement. Specifically, the inclusion of FFEM increases $\alpha_{xx}$ and decreases 
$\alpha_{zz}$ relative to the ZSISA approximation.

Comparison with literature reveals that the ZSISA results from Ref.~\cite{rostami_anisotropic_2025}—which also utilize a norm-conserving PseudoDojo pseudopotential—nearly coincide with our own ZSISA values. In contrast, the calculations by Ref.~\cite{masuki_full_2023}, which employ the PAW method, yield $\alpha_{zz}$ values slightly lower than experiment. Given that all three studies utilize the PBEsol functional, the discrepancy in $\alpha_{zz}$ may be attributed to the different pseudopotentials. Consistent with our findings, Ref.~\cite{masuki_full_2023} also observes that FFEM increases $\alpha_{xx}$ and decreases $\alpha_{zz}$ compared to ZSISA, though the magnitude of this correction is slightly smaller than in our work.

The pressure dependence of thermal expansion and the comparative impact of the FFEM are illustrated in Figs.~\ref{fig:alpha_xx} and \ref{fig:alpha_zz}, which depict $\alpha_{xx}$ and $\alpha_{zz}$ at $0$, $40$, and $80$ kbar. For $\alpha_{xx}$, the difference between ZSISA and FFEM is negligible. However, $\alpha_{zz}$ exhibits a distinct difference that persists across the pressure range, with FFEM consistently yielding lower values than ZSISA. Since our free energy is fitted using a fourth-degree polynomial across the $5\times 5$ lattice parameter grid, these detectable differences in $\alpha_{zz}$ can be attributed to higher-order terms that are not accounted for by the ZSISA theorem.

The internal thermal expansion exhibits significant discrepancies when computed using the ZSISA versus the FFEM framework. Figure~\ref{fig:ut_zno} displays the internal parameter $u$ calculated within both approximations, while the corresponding internal thermal expansion is shown in Fig.~\ref{fig:alpha_int_zno}. For comparison, we include experimental $u(T)$ values from Ref.~\cite{albertsson_atomic_1989} alongside theoretical results from Liu et al.~\cite{liu_internal_2018, liu_erratum_2018}. The results in Ref.~\cite{liu_internal_2018} were derived via free energy minimization—consistent with our approach—whereas those in Ref.~\cite{liu_erratum_2018} were obtained using analytical expressions for temperature induced change of the internal parameter $\Delta u(T)$. Within Gr\"uneisen theory, these two methods should yield equivalent results.

Our data are in reasonable agreement with the LDA-based findings of previous authors. Furthermore, our results are expected to align with Gr\"uneisen $E^{(3)}+F_{vib}^{(2)}$ theory, despite our use of a fourth-order polynomial for both the static energy and free energy, as well as a different exchange-correlation functional. Notably, we have omitted the experimental data from Ref.~\cite{timoshenko_temperature_2014}, as they suggest a significantly more rapid increase of $u$ with temperature than predicted by our models or other available literature.

The remaining thermodynamic quantities, specifically the isobaric heat capacity and the adiabatic bulk modulus, are detailed and discussed in the Supplemental Material.

In the following section, we compare the calculated temperature dependence of the elastic constants with available experimental data. Because experimental measurements are conducted under varying electrical boundary conditions—specifically at constant ${\bf E}$ (electric field) or constant ${\bf D}$ (electric displacement)—we first determine the elastic constants at a constant zero electric field (${\bf E}=0$). We then apply the thermodynamic relationships defined in Eq.~\ref{c_d_e} to derive the corresponding elastic constants at constant ${\bf D}$.

We begin with an analysis of the elastic constants calculated at a constant zero electric field (${\bf E}=0$). Table~\ref{table:2} summarizes the elastic constants at the equilibrium geometry, providing both clamped-ion and relaxed-ion values obtained via the QHA method at the lowest studied temperature ($4$ K). These values account for zero-point motion effects (ZPME) on both the free energy and the crystal geometry.

We observe substantial discrepancies between the clamped-ion and relaxed-ion results: $656$ kbar ($31
\%$) for $C_{11}$, $292$ kbar ($23 \%$) for $C_{12}$, $381$ kbar ($35 \%$) for $C_{13}$, $750$ kbar ($37 \%$) for $C_{33}$, and $178$ kbar ($45 \%$) for 
$C_{44}$. These large differences underscore the critical role of internal relaxations in determining the elastic properties of ZnO, a finding consistent with previous LDA-based studies by Wu et al.~\cite{wu_systematic_2005}.

To facilitate a comparison with experiment, we estimated the zero-temperature experimental elastic constants by taking the $300$ K values from Ref.~\cite{kobiakov_elastic_1980} and adjusting them using the theoretical temperature-dependent shifts determined in our work (and corroborated by Ref.~\cite{rostami_anisotropic_2025}). Our extrapolated $0$ K experimental values—benchmarked against data from Ref.~\cite{rostami_anisotropic_2025}-are $C_{11}=2152$ $(2132)$ kbar, $C_{12}=1230$ $(1236)$ kbar, $C_{13}=1107$ $(1103)$ kbar, $C_{33}=2153$ $(2150)$ kbar, and $C_{44}=496$ $(448)$ kbar. Using these as references, the relative errors for our PBEsol calculations are $-2.6 \%$ for $C_{11}$, $1 \%$ for 
$C_{12}$, $2 \%$ for $C_{13}$, $5 \%$ for $C_{33}$, 
and $-17 \%$ for $C_{44}$. With the exception of $C_{44}$, these deviations fall within the typical range for DFT-based predictions.

Finally, comparing our results with existing PBEsol literature shows reasonable agreement. Values from Ref.~\cite{rostami_anisotropic_2025} (extracted from their Fig.~3c) were compared to our $4$ K constant ${\bf E}$ values including ZPME. 
The resulting differences are: $\Delta C_{11}=69$ kbar ($5 \%$), $\Delta C_{12}=-6$ kbar ($-0.5 \%$), $\Delta C_{13}=43$ kbar ($4 \%$), $\Delta C_{33}=-51$ kbar ($-2 \%$), and $\Delta C_{44}=46$ kbar ($12 \%$). Since both studies utilize the same norm-conserving pseudopotentials, these minor discrepancies likely arise from different numerical methodologies (DFPT in Ref.~\cite{rostami_anisotropic_2025} versus numerical differentiation in \texttt{thermo\_pw}). Other reported PBEsol values~\cite{wang_thermodynamic_2014} exhibit significantly larger deviations. A comparison of the QSA values from this source with our calculation is given in the supplemental material. 

 \begin{table*}
  \begin{center}
\caption{PBEsol static dielectric constants ($\epsilon_{\alpha\beta}^{(0)}$), high-frequency dielectric constants ($\epsilon_{\alpha\beta}^{\infty}$), and Born effective charges ($Z^*_{s\alpha\beta}$) for ZnO. These properties were determined by interpolating the values calculated via DFPT at the configurations corresponding to the minimum of the free energy.
}%
\begin{tabular}{llllllll}
\toprule
 &  T (K)  &$\epsilon^\infty_{1}$  & $\epsilon^\infty_{3}$ & $\epsilon^{(0)}_{1}$ & $\epsilon^{(0)}_{3}$  & $Z^*_{O11}$  & $Z^*_{O33}$  \\
 This work (PBEsol)  &  $7$  &$6.03$  & $5.72$ & $11.2$ & $11.7$  & $-2.15$  & $-2.18$  \\
Ref.~\cite{wu_systematic_2005} (LDA)  &  $0$  &$5.76$  & $5.12$ & $10.31$ & $10.27$  &  &   \\
\hline
\botrule
\end{tabular}
\label{table:4}
\end{center}
\end{table*}

The temperature-dependent elastic constants (TDECs) were computed using both the QSA and QHA methods; a comparative analysis of these approximations is provided in the Supplemental Material. In this section, we focus exclusively on the QHA results evaluated at five equilibrium geometries along the stress-pressure curve and interpolated as detailed above. Figure~\ref{fig:el_cons_qha} presents our calculated QHA TDECs at $0$ kbar over a temperature range of $0$ to $800$ K. We compare our results with the temperature-dependent measurements of $C^{\bf E}_{44}={ 1 \over S^{\bf E}_{55}}$ from Ref.~\cite{kobiakov_elastic_1980}, as well as the $300$ K benchmark values from the same source and the theoretical data from Ref.~\cite{rostami_anisotropic_2025}. With the exception of the $0$ K value for $C_{11}$, our data—including the temperature-dependent trends—align closely with those in Ref.~\cite{rostami_anisotropic_2025}. A notable distinction is that we find $C_{11} > C_{33}$ by $45$ kbar, whereas Ref.~\cite{rostami_anisotropic_2025} reports $C_{33}>C_{11}$ by $18$ kbar.
\begin{figure}
\centering
\includegraphics[width=\linewidth]{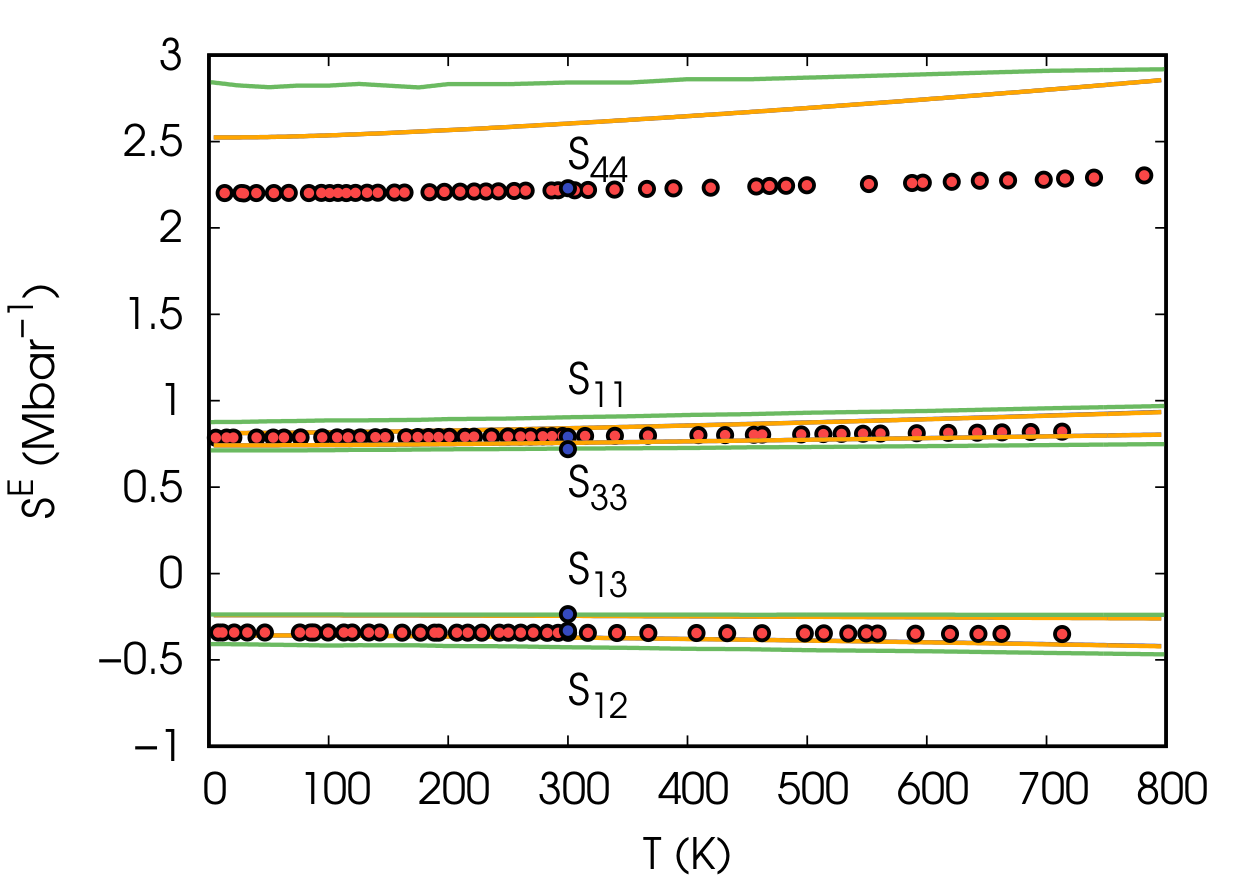}
\caption{PBEsol elastic compliances $S_{ij}$ of ZnO (calculated under constant electric field, ${\bf E}$, conditions) as a function of temperature at $0$ kbar. Orange lines represent the adiabatic elastic compliances obtained within the V-ZSISA framework by interpolating the elastic constants across five reference geometries. The green lines denote the PBEsol calculations from Ref.~\cite{rostami_anisotropic_2025}. Experimental data from Ref.~\cite{kobiakov_elastic_1980} are indicated by red (all temperatures) and blue dots
($300$ K), representing measurements taken under consistent boundary conditions.}
\label{fig:el_comp_qha}
\end{figure}

\begin{figure}
\centering
\includegraphics[width=\linewidth]{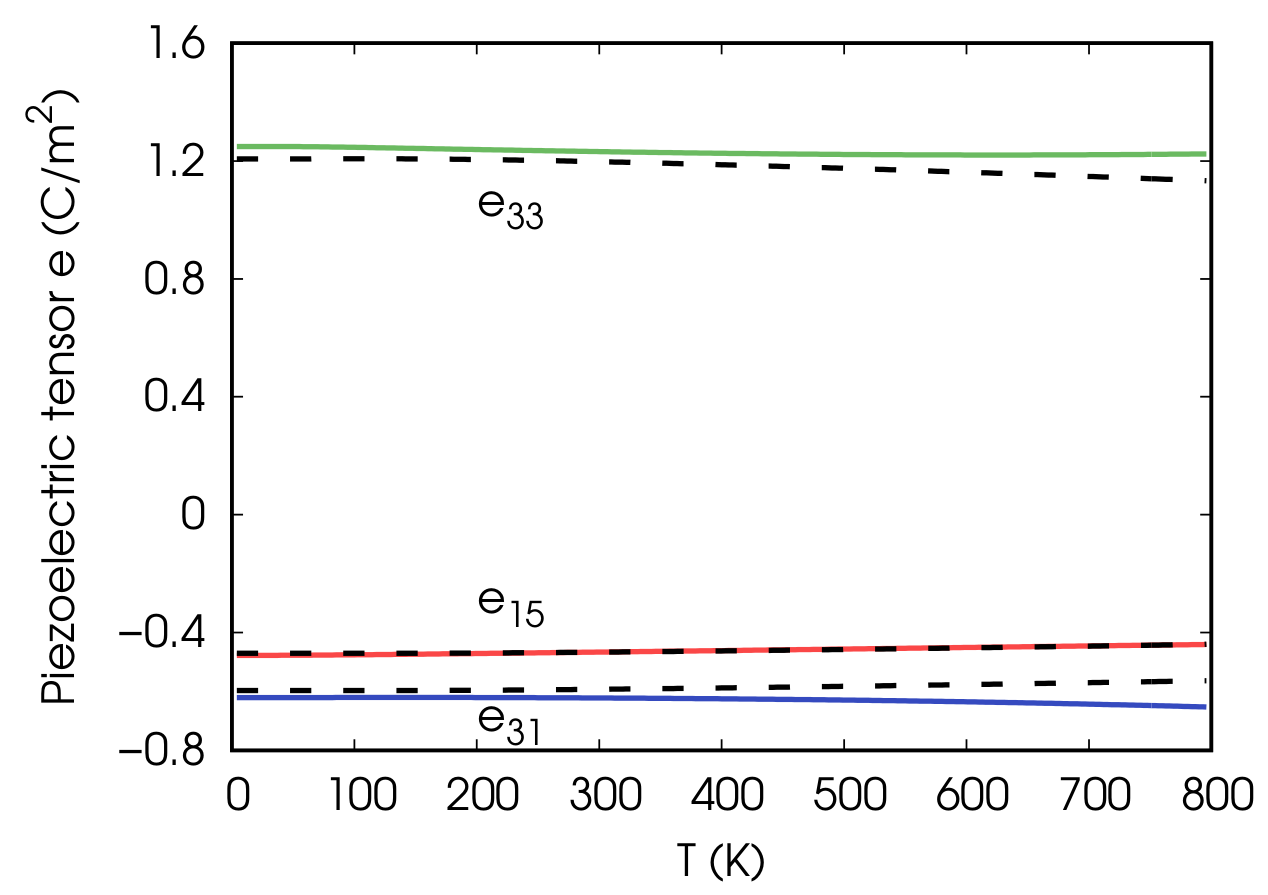}
\caption{PBEsol piezoelectric tensor components $e_{31}$, $e_{33}$, and $e_{15}$ of ZnO as a function of temperature. The solid colored lines represent results where the internal degrees of freedom are fully relaxed using the FFEM. The dashed lines correspond to the ZSISA.}
\label{fig:piezo_zsisa_ffem}
\end{figure}
Regarding the experimental comparison for $C_{44}$, while the theoretical magnitudes are lower than the measured values, the temperature dependence is reasonably well-reproduced (a more detailed analysis follows). At $300$ K, our theoretical values closely match the experimental data from Ref.~\cite{kobiakov_elastic_1980}, though our 
$C_{33}$ is slightly lower than the experimental value. Interestingly, the data from Ref.~\cite{rostami_anisotropic_2025} show $C_{11}$ as the parameter that is slightly lower than experiment.
Finally, Fig.~\ref{fig:el_cons_qha_80} illustrates the same elastic constants under a pressure of $80$ kbar. The agreement with Ref.~\cite{rostami_anisotropic_2025} remains strong, even at this elevated pressure, although we continue to find nearly degenerate values for $C_{11}$ and 
$C_{33}$.

\begin{figure}
\centering
\includegraphics[width=\linewidth]{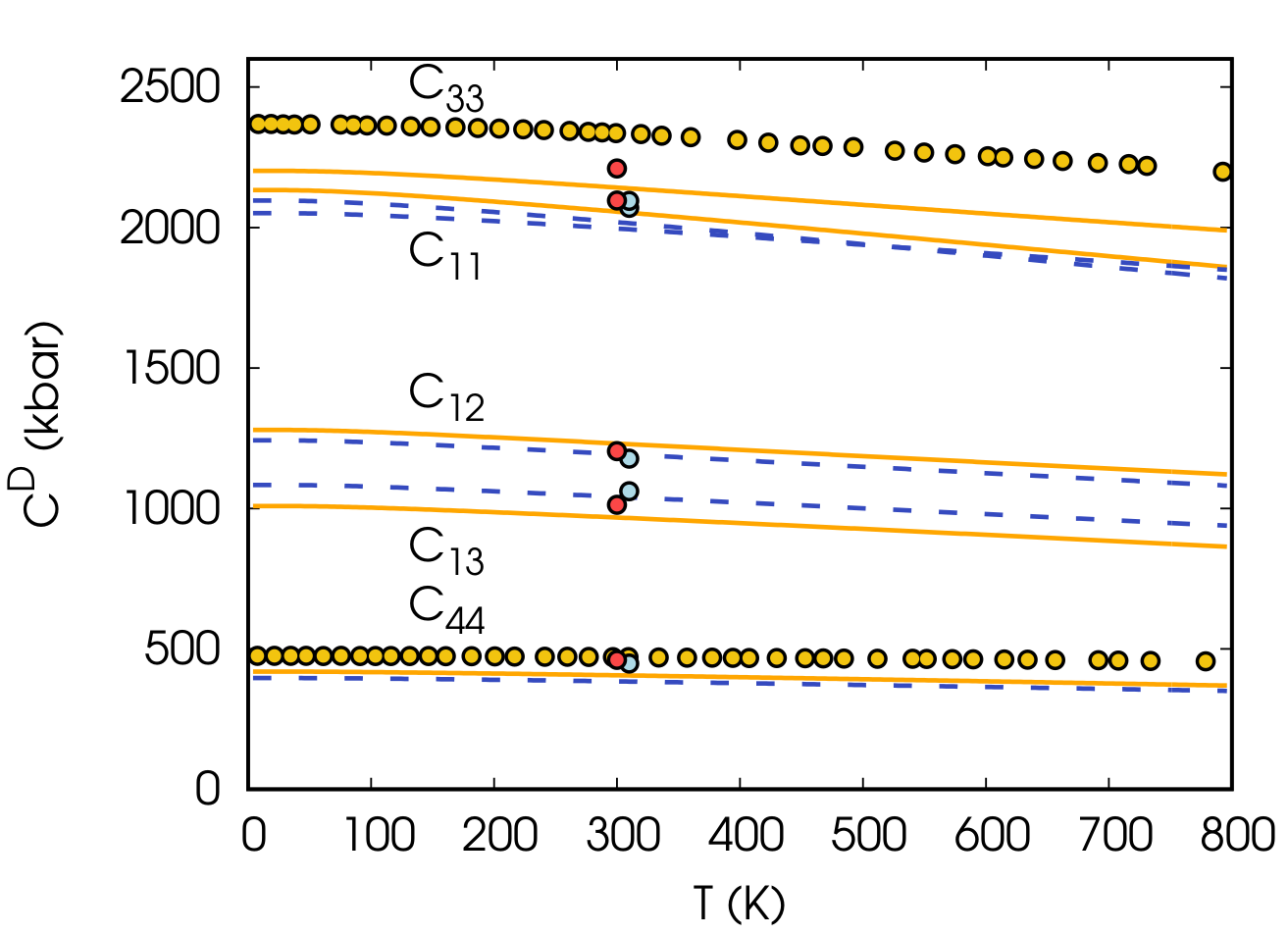}
\caption{PBEsol elastic constants of ZnO as a function of temperature at $0$ kbar, comparing different electrical boundary conditions. Orange lines represent the "stiffened" elastic constants calculated at constant ${\bf D}$, while dashed blue lines indicate the constants ${\bf E}$. Experimental data from Ref.~\cite{kobiakov_elastic_1980} are shown for constant ${\bf D}$ conditions (yellow dots). Specific $300$ K measurements are highlighted for constant ${\bf D}$ (red dots) and constant ${\bf E}$ (light blue dots). For visual clarity and to avoid overlap, the light blue dots have been offset to $310$ K.}
\label{fig:el_cons_d_e}
\end{figure}
A direct comparison with the experimental data from Ref.~\cite{kobiakov_elastic_1980} can be performed using the temperature-dependent elastic compliances 
$S_{11}$, $S_{12}$, and $S_{44}$, which were measured under constant ${\bf E}$ conditions. It should be noted that these three measured quantities alone are insufficient to invert the compliance matrix and derive the full set of elastic constants at each temperature. Figure~\ref{fig:el_comp_qha} presents our calculated compliances alongside values we derived by inverting the elastic constants reported in Ref.~\cite{rostami_anisotropic_2025}. On the scale shown, the agreement between theory and experiment is quite robust for $S_{11}$ and $S_{12}$. The primary discrepancy appears in the $S_{44}$ component, particularly regarding the $0$ K values. While our data for $S_{44}$ show a stronger temperature dependence, the results from Ref.~\cite{rostami_anisotropic_2025} exhibit a more moderate temperature sensitivity that aligns more closely with the experimental slope.

We now turn to the piezoelectric tensor and its associated temperature dependence. Several factors contribute to this dependence: first, the variation of external lattice parameters with temperature; and second, the resulting variation in the derivative of the internal coordinate $u$ with respect to strain. While an intrinsic variation of polarization at fixed crystal parameters is more complex to quantify and is beyond the scope of this work, we do account for the variation of the Born effective charges and the clamped-ion term as functions of the lattice parameters. For both quantities, we utilize the QSA framework and compare the ZSISA and FFEM approaches for the derivative of the internal coordinate with respect to strain.

Table~\ref{table:3} summarizes the values calculated at the $4$ K and $300$ K equilibrium geometries. We report both clamped-ion and relaxed-ion contributions. Within our computational scheme, the small temperature dependence of the piezoelectric tensor arises primarily from the ionic contribution. In contrast, the clamped-ion term, treated within the QSA, exhibits no detectable change in the reported digits between $4$ K and $300$ K. These results are in good agreement with previous literature values, despite differences in the exchange-correlation functionals used in those studies.

The temperature dependence of the piezoelectric tensor, calculated from $4$ to $800$ K, is illustrated in Fig.~\ref{fig:piezo_zsisa_ffem}. We observe only minor differences between the ZSISA and FFEM frameworks regarding the derivative of the internal parameter $u$ (or $x$) with respect to strain. Consequently, the piezoelectric tensor components remain nearly constant across the entire temperature range. This stability is in qualitative agreement with the nearly temperature-independent electromechanical coupling reported in Ref.~\cite{kobiakov_elastic_1980}. 

To convert the elastic constants from constant ${\bf E}$ to constant ${\bf D}$, the static dielectric constant of the solid is required. We computed the high-frequency dielectric tensor, $\epsilon^\infty_{\alpha\beta}$, using DFPT. The static dielectric constants, $\epsilon^{(0)}_{\alpha\beta}$, were then derived using Eq.~\ref{epsilon0} for the five equilibrium geometries along the stress-pressure curve. Their temperature dependence was determined by interpolating these values at the equilibrium volume corresponding to each temperature.
Table~\ref{table:4} reports the calculated high-frequency $\epsilon^{\infty}_{11}$ and $\epsilon^{\infty}_{33}$, and static dielectric constants $\epsilon^{(0)}_{11}$ and $\epsilon^{(0)}_{33}$, alongside the Born effective charges ($Z^*_{O11}$, $Z^*_{O33}$, $Z^*_{Zn11}$,
$Z^*_{Zn33}$), at the geometry corresponding to the lowest studied temperature ($4$ K).

The calculated QSA temperature dependence of the static dielectric constant is presented in the Supplemental Material and compared with the experimental data from Ref.~\cite{kobiakov_elastic_1980}. The experimental results show a rapid increase in the dielectric constants above approximately $400$ K-- a trend that is not captured by our calculations. Even within the 
$0$–$400$ K range, our model slightly underestimates the observed temperature dependence. Furthermore, the anisotropy of the dielectric constants (the difference between $\epsilon_{11}^{(0)}$ and $\epsilon_{33}^{(0)}$), even at 0 K, is significantly more pronounced in experiments than in our theoretical results, corroborating the findings of Ref.~\cite{wu_systematic_2005} that finds almost degenerate values.

As observed in other materials, the QSA framework tends to underestimate the temperature variation of the dielectric constant~\cite{giustino_electron-phonon_2017}. A more comprehensive treatment would require the inclusion of electron-phonon coupling effects; however, this approach is not currently integrated into our workflow. We reserve the implementation of these effects for future investigations.

Using the dielectric and piezoelectric quantities discussed above, we determined the elastic constants under constant ${\bf D}$ conditions. Figure~\ref{fig:el_cons_d_e} illustrates these results in comparison with their constant electric field 
(${\bf E}$) counterparts. Consistent with the predictions of Eq.~\ref{c_d_e} and the signs of the piezoelectric tensor components, all elastic constant components at constant ${\bf D}$ are higher than those at constant ${\bf E}$, with the exception of $C_{13}$. Experimentally, temperature-dependent data from Ref.~\cite{kobiakov_elastic_1980} are only available for $C_{33}$ and $C_{44}$, although the full set of elastic constants has been reported at 300 K.

Overall, the calculated temperature dependence and the magnitude of the shift between constant ${\bf D}$ and constant ${\bf E}$ states align well with the experimental scale. While some discrepancies are visible in the $0$ K absolute values, we note a lack of consistency in the experimental literature for 
$C_{33}$; specifically, the $300$ K value from the temperature-dependent series does not match the standalone $300$ K benchmarks (indicated by yellow and red dots). This inconsistency reflects the inherent experimental uncertainty in determining the absolute value of this constant. Finally, because the shift for $C_{33}$ is more pronounced than that for $C_{11}$, the constant ${\bf D}$ results yield the relation 
$C_{33}>C_{11}$.

\begin{figure}
\centering
\includegraphics[width=\linewidth]{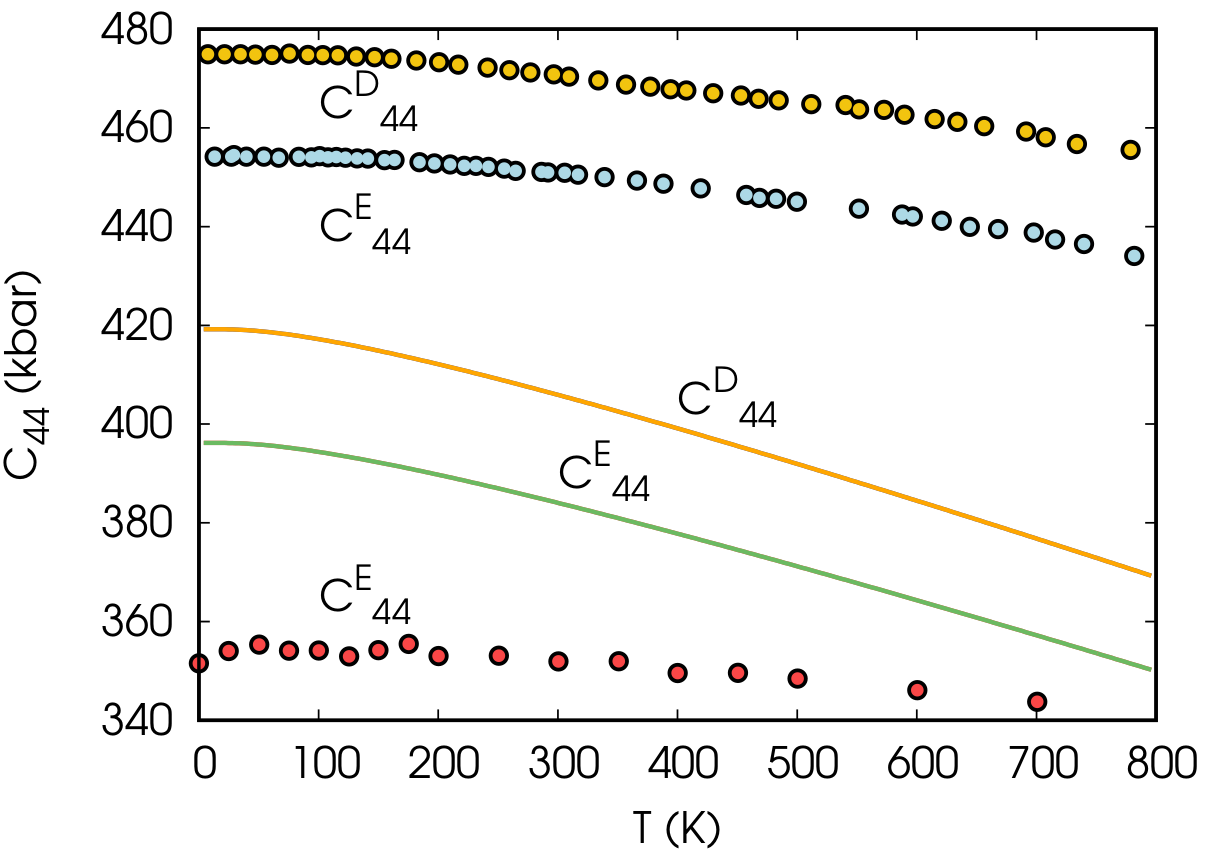}
\caption{PBEsol $C_{44}$ elastic constant of ZnO as a function of temperature at $0$ kbar, illustrating the effects of electrical boundary conditions. Calculations performed at constant electric field 
(${\bf E}$) are shown in green, while constant electric displacement (${\bf D}$) conditions are shown in orange. For $C_{44}$ adiabatic and isothermal elastic constants coincide. Red dots indicate the PBEsol results from Ref.~\cite{rostami_anisotropic_2025}. Experimental values from Ref.~\cite{kobiakov_elastic_1980} are included for both constant ${\bf D}$ (yellow dots) and constant ${\bf E}$ (light blue dots) boundary conditions, highlighting the piezoelectric stiffening of the shear mode.}
\label{fig:el_cons_c44_d_e}
\end{figure}

\begin{figure}
\centering
\includegraphics[width=\linewidth]{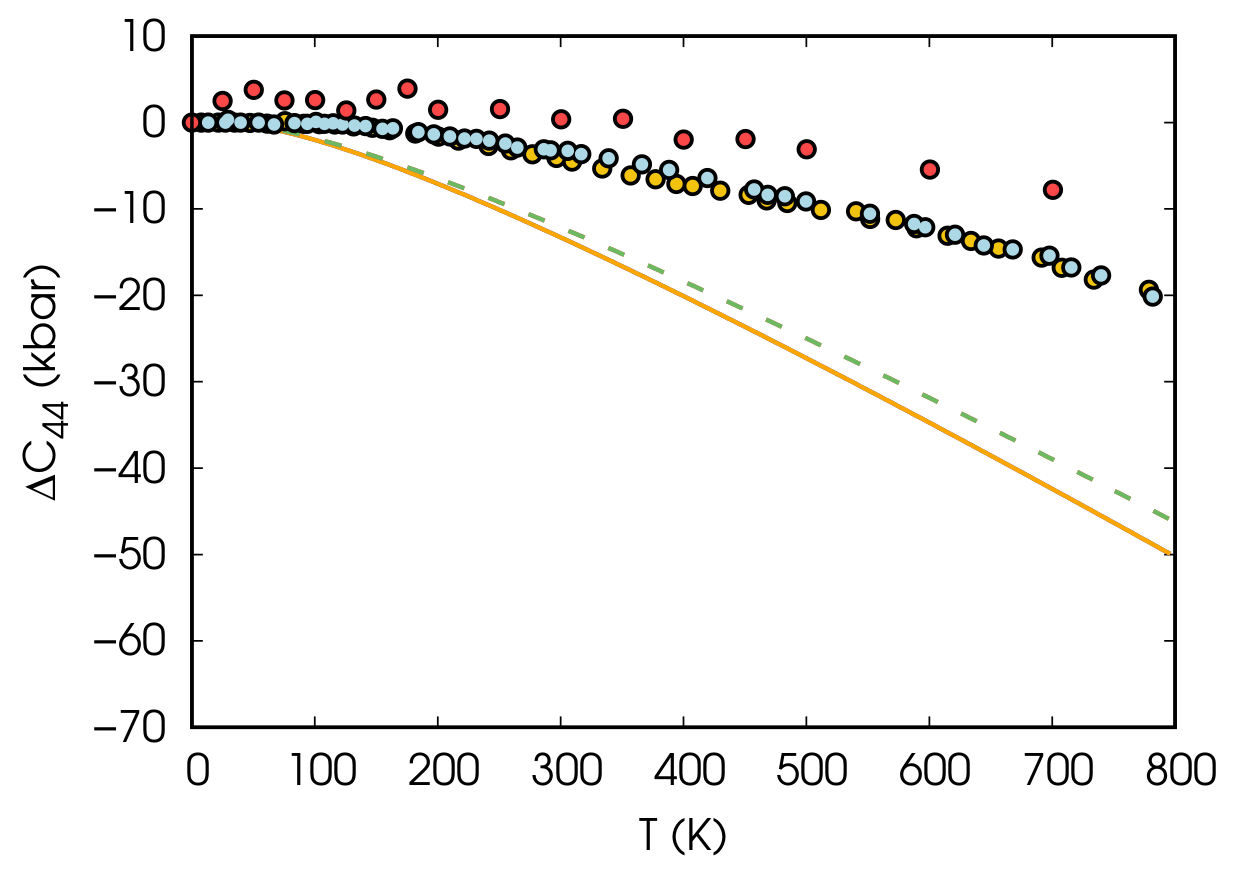}
\caption{PBEsol $C_{44}$ elastic constant of ZnO as a function of temperature at $0$ kbar. Results are shown for both constant ${\bf E}$ (dashed green) and constant ${\bf D}$ (orange) boundary conditions. For this specific shear mode, the adiabatic and isothermal elastic constants are identical. Red dots indicate the constants ${\bf E}$ PBEsol results from Ref.~\cite{rostami_anisotropic_2025}. Experimental data from Ref.~\cite{kobiakov_elastic_1980} are provided for comparison, with yellow dots representing constant ${\bf D}$ measurements and light-blue dots representing constant ${\bf E}$ measurements. All values are shown relative to their respective $0$ K magnitudes.}
\label{fig:delta_c44_d_e}
\end{figure}

\begin{figure}
\centering
\includegraphics[width=\linewidth]{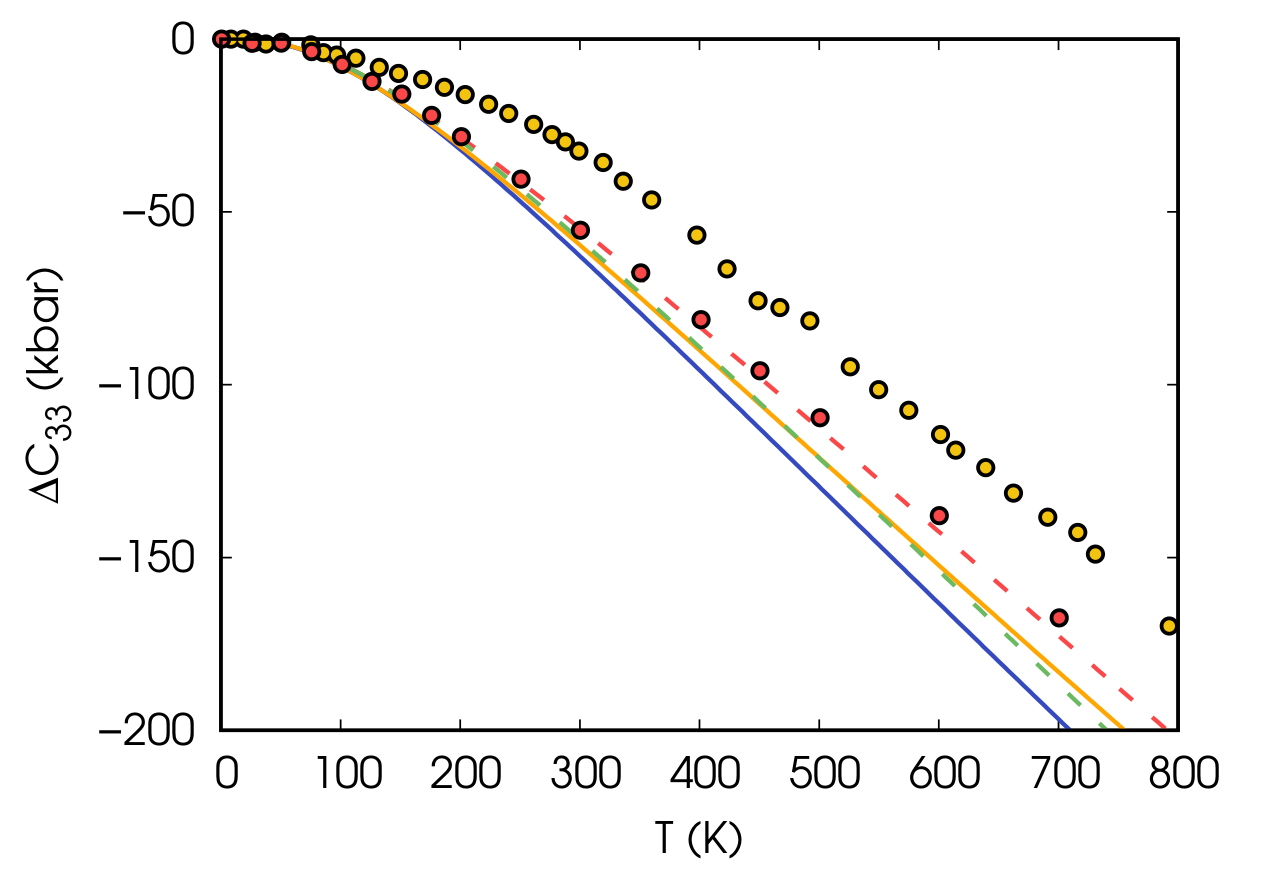}
\caption{PBEsol $C_{33}$ elastic constant of ZnO as a function of temperature at $0$ kbar. Results are shown for constant electric field (${\bf E}$) and constant electric displacement (${\bf D}$) boundary conditions. All values are shown relative to their respective $0$ K magnitudes to highlight temperature-induced variations. Isothermal elastic constants are represented by dashed green ({\bf E}) and blue (${\bf D}$) lines, while adiabatic constants are shown as dashed red ${\bf E}$ and orange (${\bf D}$) lines. For comparison, the red dots indicate the adiabatic constant ${\bf E}$ results from Ref.~\cite{rostami_anisotropic_2025} (comparable to our dashed red curve), and yellow dots denote experimental constant ${\bf D}$ measurements from Ref.~\cite{kobiakov_elastic_1980} (comparable to our orange curve).}
\label{fig:delta_c33_d_e}
\end{figure}

In Fig.~\ref{fig:el_cons_c44_d_e}, we examine the elastic constant $C_{44}$ under both constant ${\bf D}$ and constant ${\bf E}$ conditions. This specific constant is the only one for which temperature-dependent measurements under both boundary conditions are available in Ref.~\cite{kobiakov_elastic_1980}. The experimental separation between these two quantities—defined by the ratio $(e_{15})^2/\epsilon^{(0)}_{1}$—is $21$ kbar at $1$ K, $19$ kbar at $305$ K, and $22$ kbar at the maximum measured temperature ($780$ K), consistent with the weak temperature dependence for the electromechanical coupling.

Our theoretical results show a stronger overall temperature dependence than the experimental data, but the separation between the curves (the splitting) is in reasonable agreement. We calculate a separation of $23$ kbar at $0$ K, $22$ kbar at $300$ K, and $20$ kbar at $780$ K.

To examine the temperature sensitivity in greater detail, we subtracted the $0$ K baseline values from both the theoretical and experimental datasets. This comparison is presented in Fig.~\ref{fig:delta_c44_d_e}. Experimentally, there is a negligible difference between the temperature-dependent slopes of $C^{\bf E}_{44}$ and $C^{\bf D}_{44}$. Our results similarly predict only a minor difference. However, our calculation overestimates the rate of softening with temperature; interestingly, this trend is not observed in the constant-${\bf E}$ data from Ref.~\cite{rostami_anisotropic_2025}, where the temperature dependence is weaker than experiment but ultimately closer than our own. Whether the ZSISA approximation contributes to this discrepancy remains a subject for further investigation.

A similar analysis for the $C_{33}$ elastic constant is presented in Fig.~\ref{fig:delta_c33_d_e}, where we plot the temperature dependence for both constant ${\bf D}$ (solid line) and constant ${\bf E}$ (dashed line) conditions.
We find excellent agreement with the temperature-dependent trends predicted (at constant ${\bf E}$) in Ref.~\cite{rostami_anisotropic_2025}. Furthermore, for this specific elastic constant, the QHA approach reasonably predicts the experimental temperature dependence. 

\begin{figure}
\centering
\includegraphics[width=\linewidth]{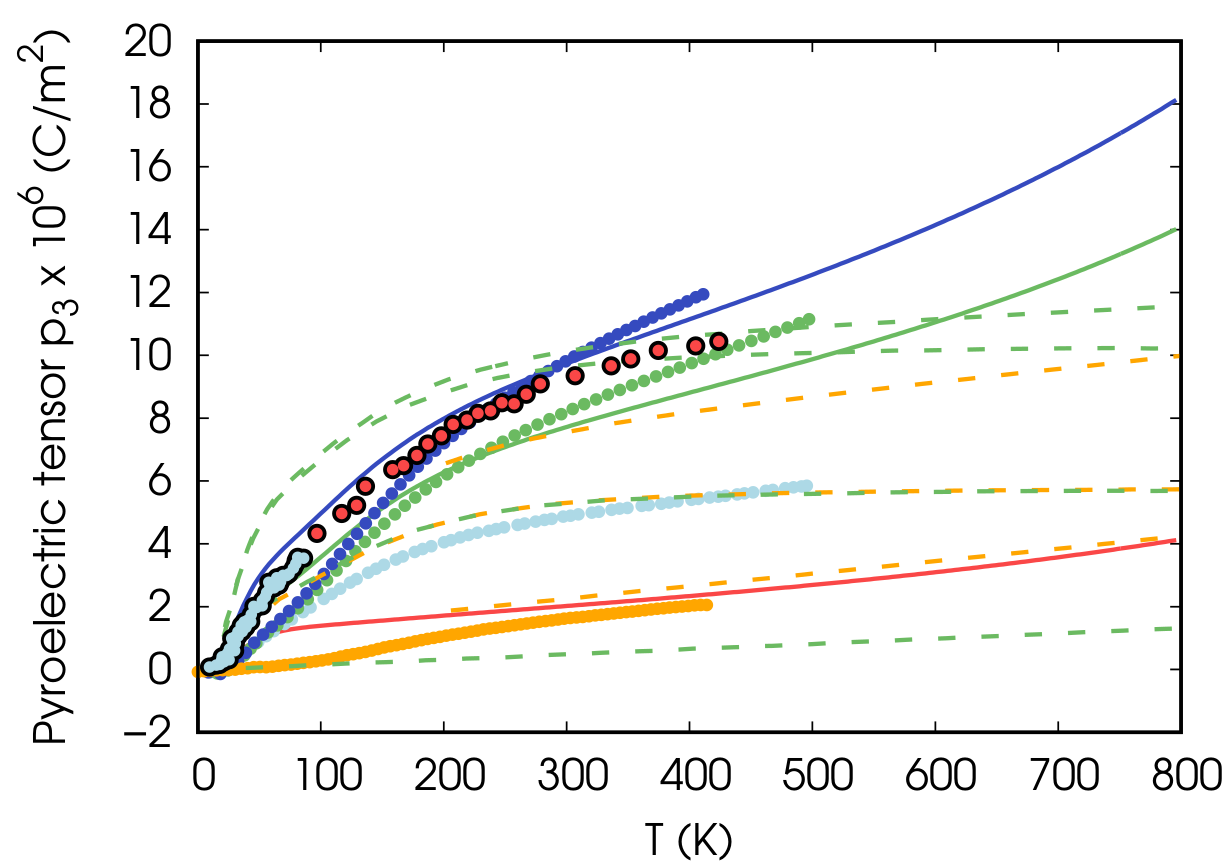}
\caption{
PBEsol pyroelectric coefficient $p_3$ of ZnO as a function of temperature at $0$ kbar. The total pyroelectric coefficient (combined primary and secondary effects) is shown as a solid blue curve, compared with experimental data from Ref.~\cite{heiland_pyroelectricity_1966} (red and light-blue large dots). The decomposition of our results shows the secondary contribution due to the total piezoelectric effect (red curve) and the extracted primary effect (green curve). For comparison, we include total pyroelectric results from Ref.~\cite{masuki_full_2023} (blue dots), Ref.~\cite{liu_first-principles_2016} (top dashed orange line), and Ref.~\cite{liu_mechanisms_2018} (top dashed green line). Further decompositions from literature are provided: dashed orange lines (from bottom to top) represent the secondary, primary, and total effects from Ref.~\cite{liu_first-principles_2016}; dashed green lines (from bottom to top) show the secondary, primary, and total contributions from Ref.~\cite{liu_mechanisms_2018}. Small dots represent the decomposition from Ref.~\cite{masuki_full_2023}: secondary/piezoelectric (orange), primary (light-blue), primary plus Born (green), and total (blue).}
\label{fig:pyro}
\end{figure}

Finally, we utilized the temperature-dependent internal parameter $u$, calculated via the ZSISA and FFEM methods, to predict the temperature dependence of the pyroelectric tensor. Figure~\ref{fig:pyro} compares our calculated pyroelectric coefficients with experimental data and existing literature. Our results are in reasonable agreement with experimental values and align well with the data reported in Ref.~\cite{masuki_full_2023} and Ref.~\cite{liu_mechanisms_2018}.

The total pyroelectric tensor is decomposed into primary and secondary contributions. The primary contribution was derived as the product of the Born effective charges and the internal thermal expansion (scaled by $u(T)$) within the FFEM framework (Fig.~\ref{fig:alpha_int_zno}). From this, we subtracted the ZSISA contribution, which is inherently included in the secondary effect (refer to Eq.~\ref{pyro1}). The secondary contribution, which arises from the coupling of the piezoelectric effect and the external thermal expansion (see Eq.~\ref{pyro1_wurt}), shows strong agreement with the findings of Liu et al.~\cite{liu_first-principles_2016}. It also matches the contribution we extracted from Ref.~\cite{masuki_full_2023}, determined by the difference between the blue and green data points in their analysis.

\section{\label{sec:level5}Conclusions}

We have presented a generalized approach for calculating the thermodynamic properties of systems with internal degrees of freedom, incorporating both the ZSISA and FFEM approximations. This method extends a previous scheme developed for hcp metals to now accommodate piezoelectric and/or pyroelectric insulators. By applying this framework to wurtzite ZnO, we analyzed the influence of internal degrees of freedom on thermal expansion, as well as the piezoelectric and pyroelectric tensors. We noted that while the ZSISA approximation allows for automated application, identifying the minimal set of internal coordinates for FFEM remains a case-specific challenge, particularly in distorted structures.

Our results demonstrate strong agreement with existing computational and experimental data. Specifically, we characterized the temperature-dependent elastic constants and compliances of ZnO under both constant ${\bf E}$ and constant ${\bf D}$ boundary conditions. Given that many components of these tensors remain experimentally uncharacterized as a function of temperature, we hope our theoretical predictions will provide the necessary impetus for further experimental validation.

While this work does not include QHA TDEC calculations within the FFEM framework, our preliminary analysis of clamped ions suggests that internal relaxations significantly impact elastic constant values. We anticipate that accounting for these effects in future studies—despite the high computational cost of 270 phonon dispersion calculations per equilibrium geometry—will further refine the agreement between theory and experiment.

Currently, our approach is feasible for systems where the subspace of internal degrees of freedom is small (dimension lower than 2 or 3), as the grid-point scaling remains exponential. Future investigations will explore methods to reduce the number of required grid points—for instance, by treating the vibrational free energy as a low-degree polynomial as in the Gr\"uneisen approach—to enhance computational efficiency. Finally, all methods described herein have been integrated into the publicly available \texttt{thermo\_pw} software package.~\cite{dal_corso_thermo_pw_2022}

\begin{acknowledgments}
Computational facilities have been provided by SISSA through its Linux Cluster, ITCS, and the SISSA-CINECA 2021-2025 Agreement.
This work has been partially supported by the Italian MUR (Ministry of University and Research) through the National Centre for HPC, Big Data, and Quantum Computing (grant No. CN00000013) and by the European Union through the MAX ``MAterials design at the eXascale" Centre of Excellence for Supercomputing applications (Grant agreement No. 101093374, co-funded by the European High Performance Computing joint Undertaking (JU) and participating countries 824143).
\end{acknowledgments}


\bibliography{apssamp}

\end{document}